\input harvmac 
\input epsf.tex

\overfullrule=0mm

\newcount\figno
\figno=0
\def\fig#1#2#3{
\par\begingroup\parindent=0pt\leftskip=1cm\rightskip=1cm\parindent=0pt
\baselineskip=11pt
\global\advance\figno by 1
\midinsert
\epsfxsize=#3
\centerline{\epsfbox{#2}}
\vskip 12pt
{\bf Fig. \the\figno:} #1\par
\endinsert\endgroup\par
}
\def\figlabel#1{\xdef#1{\the\figno}}
\def\encadremath#1{\vbox{\hrule\hbox{\vrule\kern8pt\vbox{\kern8pt
\hbox{$\displaystyle #1$}\kern8pt}
\kern8pt\vrule}\hrule}}

\def\wrt{with respect to\ }
\def\rhs{right hand side\ }

\def\IR{\relax{\rm I\kern-.18em R}}
\font\cmss=cmss10 \font\cmsss=cmss10 at 7pt

\font\cmss=cmss10 \font\cmsss=cmss10 at 7pt
\def\IZ{\relax\ifmmode\mathchoice
{\hbox{\cmss Z\kern-.4em Z}}{\hbox{\cmss Z\kern-.4em Z}}
{\lower.9pt\hbox{\cmsss Z\kern-.4em Z}}
{\lower1.2pt\hbox{\cmsss Z\kern-.4em Z}}\else{\cmss Z\kern-.4em Z}\fi}
\def\IN{\relax{\rm I\kern-.18em N}}


\Title{\vbox{\hsize=3.truecm \hbox{SPhT/02-096}}}
{{\vbox {
\centerline{Rectangular Matrix Models}
\bigskip
\centerline{and Combinatorics of Colored Graphs}
}}}
\bigskip
\centerline{ 
P. Di Francesco\foot{philippe@spht.saclay.cea.fr} 
}
\medskip
\centerline{ \it CEA-Saclay, Service de Physique Th\'eorique,}
\centerline{ \it F-91191 Gif sur Yvette Cedex, France}
\bigskip
\noindent 
We present applications of rectangular matrix models to various
combinatorial problems, among which the enumeration of face-bicolored
graphs with prescribed vertex degrees, and vertex-tricolored triangulations. 
We also mention possible applications to Interaction-Round-a-Face and 
hard-particle statistical models defined on random lattices.

\Date{08/02}

\nref\BIPZ{E. Br\'ezin, C. Itzykson, G. Parisi and J.-B. Zuber, {\it Planar
Diagrams}, Comm. Math. Phys. {\bf 59} (1978) 35-51.}
\nref\DGZ{P. Di Francesco, P. Ginsparg
and J. Zinn--Justin, {\it 2D Gravity and Random Matrices},
Physics Reports {\bf 254} (1995) 1-131.}
\nref\PDF{P. Di Francesco, {\it Matrix Model Combinatorics: Applications to Folding and
Coloring}, in {\it Random Matrix Models and their
Applications}, Bleher and Its Eds., MSRI Publications Vol. {\bf 40} 111-170,
Cambridge University Press (2001).} 
\nref\EY{B. Eynard, {\it Random Matrices}, Saclay Lecture Notes (2000),
available at {\sl http://www-spht.cea.fr/lectures\_notes.shtml} }
\nref\TUT{W. Tutte,
{\it A Census of planar triangulations}
Canad. Jour. of Math. {\bf 14} (1962) 21-38;
{\it A Census of Hamiltonian polygons}
Canad. Jour. of Math. {\bf 14} (1962) 402-417;
{\it A Census of slicings}
Canad. Jour. of Math. {\bf 14} (1962) 708-722;
{\it A Census of Planar Maps}, Canad. Jour. of Math.
{\bf 15} (1963) 249-271.}
\nref\SCH{G. Schaeffer,  {\it Bijective census and random
generation of Eulerian planar maps}, Electronic
Journal of Combinatorics, vol. {\bf 4} (1997) R20 and {\it Conjugaison d'arbres
et cartes combinatoires al\'eatoires} PhD Thesis, Universit\'e
Bordeaux I (1998).}
\nref\BMS{M. Bousquet-M\'elou and G. Schaeffer,
{\it Enumeration of planar constellations}, Adv. in Applied Math.,
{\bf 24} (2000) 337-368.}
\nref\PS{D. Poulhalon and G. Schaeffer,
{\it A note on bipartite Eulerian planar maps}, preprint (2002),
available at {\sl http://www.loria.fr/$\sim$schaeffe/}}
\nref\BDE{J. Bouttier, P. Di Francesco and E. Guitter, 
{\it Counting colored random triangulations}, preprint t02/075 
and cond-mat/0206452, to appear in Nucl. Phys. {\bf B} (2002).}
\nref\MOR{T. Morris, {\it Chequered surfaces and complex matrices}, 
Nucl. Phys. {\bf B356} (1991) 703-728.}
\nref\BON{B. Bonk, J. Math. Phys. {\bf 6} (1965) 228.}
\nref\AMP{A. Anderson, R. Myers and V. Periwal, Phys. Lett. {\bf B 254} (1991) 89-93;
Nucl. Phys. {\bf B 360} (1991) 463-479. See also
R. Myers and V. Periwal, {\it From polymers to quantum gravity:
triple-scaling in rectangular random matrix models}, Nucl. Phys. {\bf B390} (1993) 716, and
R. Lafrance and R. Myers, {\it Flows for rectangular matrix models},
Mod. Phys. Lett. {\bf A9} (1994) 101-113.}
\nref\VER{J. Verbaarschot and I. Zahed, {\it Spectral density of the QCD Dirac operator
near zero virtuality},  Phys. Rev. Lett. {\bf 70} (1993) 3852-3855.}
\nref\FOR{P. Forrester, {\it The spectrum edge of random matrix ensembles},
Nucl. Phys. {\bf B402[FS]} (1993) 709-728 and {\it Exact results and universal 
asymptotics in the Laguerre matrix ensemble}, J. Math. Phys. {\bf 35} (1994) 2539-2551.}
\nref\NAG{T. Nagao and K. Slevin, {\it Nonuniversal correlation functions for random matrix
ensembles}, J. Math. Phys. {\bf 34} (1993), 2075-2085.}
\nref\BOF{A. Borodin and P. Forrester, {\it Increasing subsequences and the hard-to-soft edge
transition in matrix ensembles}, preprint math-ph/0205007 (2002).}
\nref\DEG{P. Di Francesco, B. Eynard and E. Guitter,
{\it Coloring Random Triangulations},
Nucl. Phys. {\bf B516 [FS]} (1998) 543-587.}
\nref\BDEnew{J. Bouttier, P. Di Francesco and E. Guitter,
{\it Census of Planar Maps: From the One-Matrix Model Solution  
to a Combinatorial Proof}, Saclay preprint t02/093 and cond-mat/0207682 (2002).}
\nref\CHAR{J. Ambjorn, C. Kristjansen and Yu. Makeenko, 
{\it Generalized Penner model to all genera}, Phys. Rev. {\bf D50} (1994) 5193-5203.}
\nref\BAT{H. Bateman, {\it Higher Transcendental Functions}, Vol. II,
McGraw-Hill (1953).}
\nref\ITZIZUB{C. Itzykson and J.-B. Zuber, {\it Matrix Integration and Combinatorics of the 
Modular Groups}, Commun. Math. Phys. {\bf 134} (1990) 197-207.}
\nref\HCIZ{M. Harish-Chandra, {\it Differential Operators on a Semi-simple Lie Algebra},
Am. Jour. Math. {\bf 79} (1957) 87-120;
C. Itzykson and J.-B. Zuber, {\it The planar approximation II}, 
J. Math. Phys. {\bf 21} (1980) 411-421.}
\nref\BROzub{R. Brower, P. Rossi and C. Tan,{\it The external field problem for QCD}, 
Nucl. Phys. {\bf B190[FS3]} (1981) 699-718.}
\nref\VERzub{A. Jackson, M. \c Sener and J. Verbaarschot, {\it Finite volume
partition functions and Itzykson-Zuber integrals}, Phys.Lett. {\bf B387} (1996) 355-360.}
\nref\WADzub{T. Akuzawa and M. Wadati, {\it Effective QCD Partition Function 
in Sectors with Non-Zero Topological Charge and Itzykson-Zuber Type Integral},
J. Phys. Soc. Jap. {\bf 67} (1998) 2151-2154.}
\nref\IK{I. Kostov,{\it Strings with discrete target space}, Nucl. Phys. {\bf B376}
(1992) 539-598.}
\nref\KKO{I. Kostov,{\it Gauge invariant matrix models for A-D-E closed strings}, 
Phys. Lett. {\bf B297} (1992) 74-81.}
\nref\BAXHH{R. J. Baxter, {\it Hard Hexagons: Exact Solution}, J. Phys. {\bf A 13}
(1980) L61-L70; R. J. Baxter and S.K. Tsang, {\it Entropy of Hard Hexagons},
J. Phys. {\bf A 13} (1980) 1023-1030.}
\nref\BAX{R. J. Baxter, {\it Planar Lattice Gases with Nearest-neighbour
Exclusion}, Annals of Combin. No. {\bf 3} (1999) 191-203 preprint cond-mat/9811264.}
\nref\BAXHS{R. J. Baxter, I. G. Enting and S.K. Tsang, {\it Hard Square Lattice
Gas}, J. Stat. Phys. {\bf 22} (1980) 465-489.}
\nref\BAXBOOK{R. J. Baxter, {\it Exactly Solved Models in Statistical Mechanics},
Academic Press, London (1984).}
\nref\BDFG{J. Bouttier, P. Di Francesco and E. Guitter,{\it Critical and Tricritical Hard objects
on Bicolourable Random Lattices: Exact Solutions}, J. Phys. A: Math. Gen. {\bf 35} (2002) 3821-3854.}
\nref\BC{E. Bender and E. Canfield, {\it The number of degree-restricted
rooted maps on the sphere}, SIAM J. of Discrete Math. {\bf 7} (1994) 9-15.}

\newsec{Introduction}

Matrix models have been extensively applied to various
domains of physics and mathematics. 
Matrix integrals may primarily be viewed as tools for generating
possibly decorated graphs of given topology \BIPZ\ (see also e.g.
the reviews \DGZ\ \PDF\ and \EY). The physical interpretation
of these decorated graphs allows to view them as configurations
of some statistical models with fluctuating ``space", namely
defined on random lattices. This framework has provided a host of toy models
for discretized quantum gravity. Indeed, physicists have engineered
matrix integrals to suit their needs: the particular choice
of integrand sets the peculiarities of each model, possibly
including interactions between site variables living on the graphs.
One then asks questions regarding thermodynamic and critical properties
of large graphs, and much work has been done toward the
identification of critical universality classes for the various models
at hand. 

Graph combinatorics has been a subject {\it per se} ever since
the ground-breaking work of Tutte, regarding the enumeration 
of various classes of planar graphs \TUT. The remarkable feature of most of
these results is the existence of algebraic functional relations for 
the generating functions involved.
When physicists started using matrix models to generate planar graphs,
some of these results and relations were immediately identified through the
algebraic character of the matrix model solutions. 
Somehow the simplicity of Tutte's results bore a ``matrix model flavor".
These methods always involved computing integrals over $N\times N$ Hermitian matrices
in the large $N$ limit, and the main trick consisted in translating the set of integration
rules into a diagrammatic technique, eventually leading to planar graphs.
On a different front,
some of these results have recently been revisited \SCH\ in the light of
an almost generic relation between this class of problems and 
that, much simpler, of enumeration of (possibly decorated) trees. 
By engineering
suitable bijections, this allowed to reinterpret many of these
functional relations in terms of trees, and to extend some of them
so as to include extra colorings [\xref\BMS-\xref\BDE].  
It was first noticed by Morris \MOR\ that {\it square complex} matrix integrals
could be used to generate so-called ``chequered" surfaces, namely graphs with
bicolorable faces, in such a way that no two faces sharing an edge be of 
the same color.
More generally, we will show that the natural tools for enumerating {\it face-colored}
graphs with prescribed numbers of faces of each color
are integrals over {\it rectangular} complex matrices.

Rectangular matrix models were first introduced in the context of
nuclear physics by Bonk \BON\ and first reemerged 
in an attempt to reach new critical models of two-dimensional
quantum gravity \AMP. More recently, rectangular matrix models were used
to calculate the partition function for effective Quantum Chromodynamics with
arbitrary numbers of flavors and quark masses \VER. 
Rectangular matrix ensembles have also been studied extensively in the
mathematics literature, where they bear the name of complex Wishart ensemble,
or Laguerre ensemble \FOR\ \NAG, and recently found some combinatorial application   
to the study of increasing subsequences in random permutations \BOF. 
The combinatorial
application we have in mind in the present paper is of a very different nature.
In Ref. \AMP, 
the graphical rules for matrix integration were shown to generate polymer-like
discrete surfaces. 
Here we will see that these rules naturally lead to discretized surfaces in the form
of face-colored random graphs of fixed topology.
Our aim is to apply known results from (rectangular) matrix integration to the
actual computation of various counting functions involving face-colored graphs. 

The paper is organized as follows. In Sect. 2
we first recall the link between Hermitian matrix integrals and graph combinatorics, 
and extend it to the case of rectangular complex matrices in relation to colored-graph
combinatorics.   In Sect. 3, we recall
a few facts on rectangular matrix integration, including the reduction to eigenvalues, various
methods of solution and give a first combinatorial application of the 
Gaussian rectangular matrix integral to the enumeration of face-bicolored graphs
of fixed genus with only one vertex. 
In Sect. 4, we present the computation of the general one-rectangular-matrix integral.
We first obtain the all-genus free energy using orthogonal polynomials, 
and then go on to the direct derivation of the planar limit, 
using a saddle-point technique.  
Sect. 5 is devoted to the problem of enumeration of vertex-tricolored random triangulations
via a rectangular multi-matrix integral. We show that the problem reduces to a 
single rectangular matrix integral with a logarithmic potential, 
and recover the result first found in \DEG\ using
a two-Hermitian matrix model and then rederived in a purely combinatorial manner in \BDE.
Section 6 gathers a few concluding remarks and two possible further applications
of rectangular matrix integrals to statistical models on 
random tessellations, namely Interaction-Round-a-Face (IRF) models and hard-particle models.

\newsec{Matrix Model Combinatorics}

\fig{Pictorial representation of Gaussian Hermitian matrix integration. The matrix elements
$M_{ij}$ are represented by oriented double half-edges (a), whose lines carry the matrix
indices $i,j$ respectively. A typical vertex representing Tr$(M^4)$ is represented in (b):
matrix indices are conserved along the oriented lines. The result of the Gaussian integration
is to connect all the half-edges into pairs forming edges (c) along the lines of which
the matrix indices are conserved.}{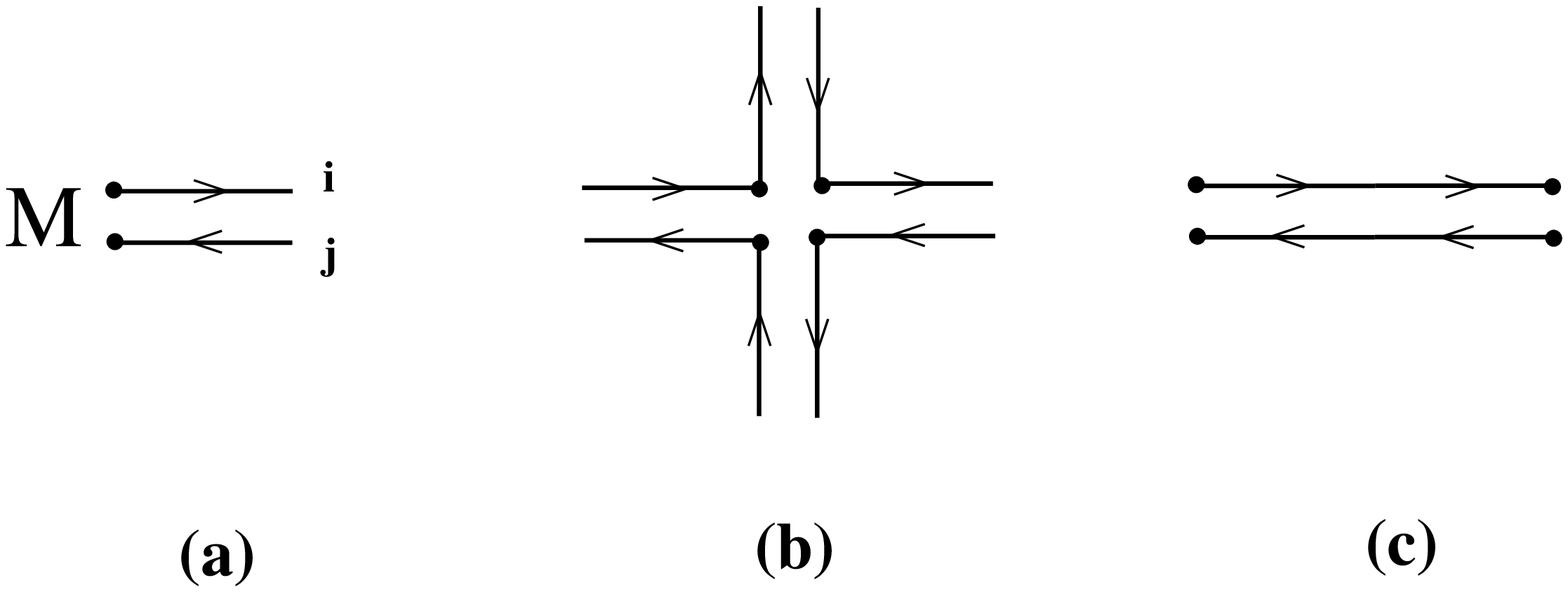}{9.cm}
\figlabel\combmat

\fig{A typical planar graph obtained from the one Hermitian matrix model integral.
The pairs of lines bordering the edges are oriented and
carry matrix indices, which must be summed over from $1$ to $N$,
resulting in a weight $N$ per face of the graph.}{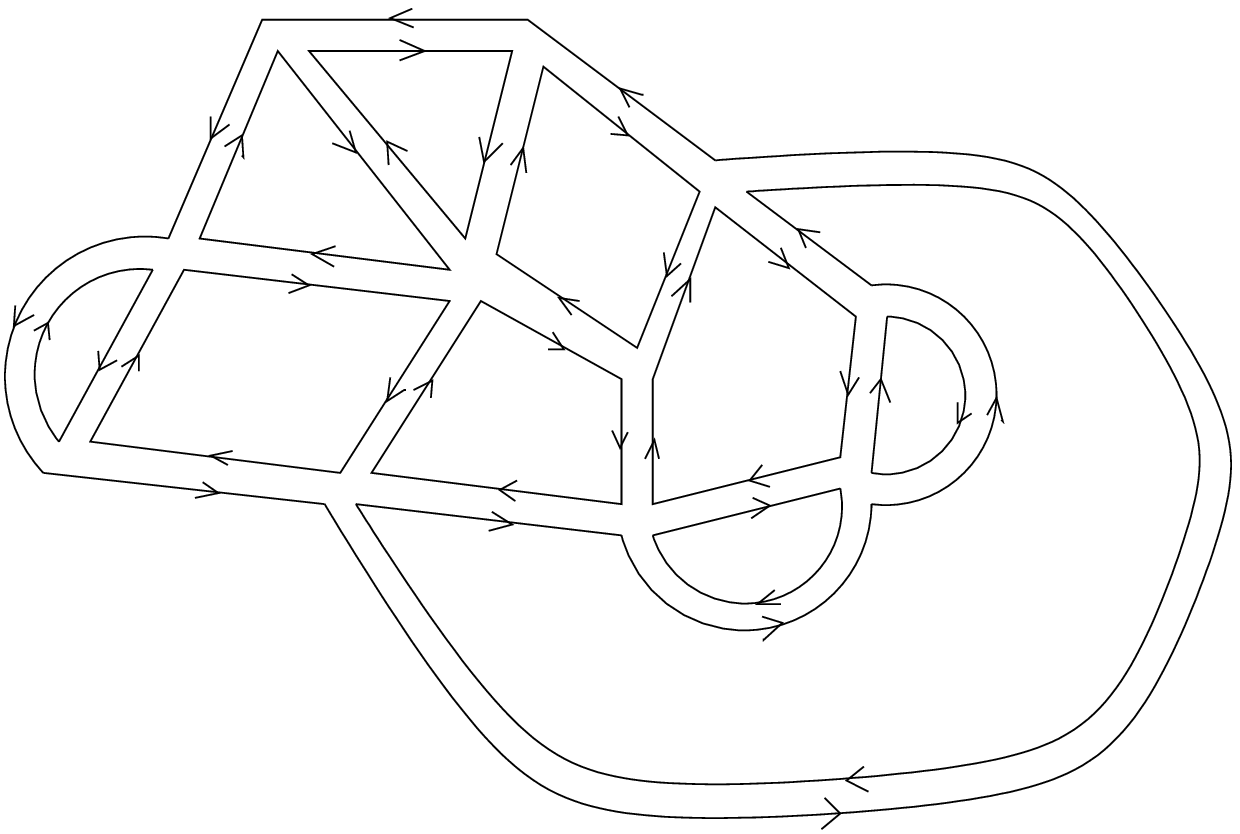}{8.cm}
\figlabel\fatso
 
As a preliminary example\foot{We refer the reader to the lecture notes \PDF\
for a more pedagogical and thorough introduction to the subject of matrix model
combinatorics.}, let us consider the $N\times N$ one Hermitian matrix integral
\eqn\hermat{ Z={\int dM e^{-N{\rm Tr}({M^2\over 2 t}-\sum_{k\geq 1} g_k {M^k\over k})}
\over \int dM e^{-N{\rm Tr}({M^2\over 2 t})}}  }
where the measure of integration is the standard flat Haar measure $dM=\prod_{i<j} dM_{ij} 
d{\bar M}_{ij} \prod_i dM_{ii}$, for which the denominator of \hermat\
plays the role of normalization factor. This ratio of integrals  
may be computed by formally expanding the $g_k$-terms
of the exponential, and by explicitly performing the term-by-term Gaussian
integration. The Gaussian integrals are alternatively calculated
by the following graphical procedure.  To each matrix element
$M_{ij}$ one associates an oriented double half-edge as shown in Fig. \combmat\ (a),
where each line carries an index of the matrix entry. In this representation, a
factor of the form ${\rm Tr}(M^k)$ is pictured as a $k$-valent vertex
with $k$ outgoing half-double edges as displayed in Fig. \combmat\ (b) for $k=4$; these
vertices carry the weight $Ng_k$. By virtue of Wick's theorem,
the result of the Gaussian integration is obtained by summing
over all possible ways of connecting the half-edges of the
integrand into pairs so as to form a closed graph; the corresponding propagator (gluing factor)
$\langle M_{ij}M_{kl}\rangle=\delta_{il}\delta_{jk}t/N$
is displayed in Fig. \combmat\ (c), and is accompanied with a weight inverse of the
prefactor of the Gaussian term in \hermat, namely $t/N$. 
Note that on the resulting graph (see Fig. \fatso\ for an illustration) 
the oriented lines forming the edges  
now carry an index $i=1,2,...,N$ of the matrix, which must be summed over in the end, resulting in 
a weight $N$ per loop of the graph, i.e. equivalently a weight $N$ per face of the graph.
The computation of \hermat\ therefore reduces to
drawing all possible fatgraphs with oriented double-lined edges like that of Fig. \fatso, 
and counting them with
a weight $Ng_k$ per $k$-valent vertex, $k=1,2,...$, as well as $t/N$ per edge 
and $N$ per face, together with an overall inverse symmetry factor coming from the various 
combinatorial numbers when expanding the exponential and resumming contributions of equivalent graphs. 
Moreover, taking the logarithm of $Z$ restricts this sum to only connected graphs,
receiving therefore an overall contribution $N^{2-2h}$, where $h$ is the genus
of the graph. Hence expanding the free energy ${\rm Log}\, Z$ at large $N$ yields a formal
power series generating arbitrary graphs with fixed topology and valences.
The graph of Fig. \fatso\ clearly has $h=0$ and contributes to the planar free energy.

The reduction of the integral \hermat\ to one over eigenvalues,
together with standard orthogonal polynomial techniques allow for a complete although tedious
calculation for all $N$. For combinatorial purposes
however it is interesting to concentrate on planar (genus zero) graphs, selected by taking
the limit $N\to \infty $. In this limit, the integral \hermat\ can be computed at leading
order by use of a saddle-point approximation, and allows in particular to recover
the generating function for rooted Eulerian maps (graphs of even valences,
namely $g_{2i-1}=0$ and $g_{2i}\neq 0$ in \hermat, and with a marked edge), 
first obtained by Tutte \TUT. 
The computation of the planar limit of \hermat\ in the case $g_{2i-1}=0$ for all $i$
is detailed in appendix A for completeness.
Note that a complete combinatorial picture involving trees
was recently found in \BDEnew\ for the general case of arbitrary $g_i$'s, including odd indices. 

\fig{Pictorial representation for Gaussian rectangular matrix integration.
The matrix elements $A_{ij},A^\dagger_{ij}$ are represented (a) as double half-edges
with oriented index lines marked $1,2$ for the corresponding range of indices $N_1,N_2$.
An extra overall orientation distinguishes between $A$ and $A^\dagger$. In (b) we have represented
a typical vertex for Tr$(AA^\dagger)^2$: the oriented lines
carry the corresponding conserved indices. The result of the Gaussian integration is to
connect the half-edges $A$ to the $A^\dagger$
into pairs through edges of the form (c); the indices are conserved along the 
oriented lines.}{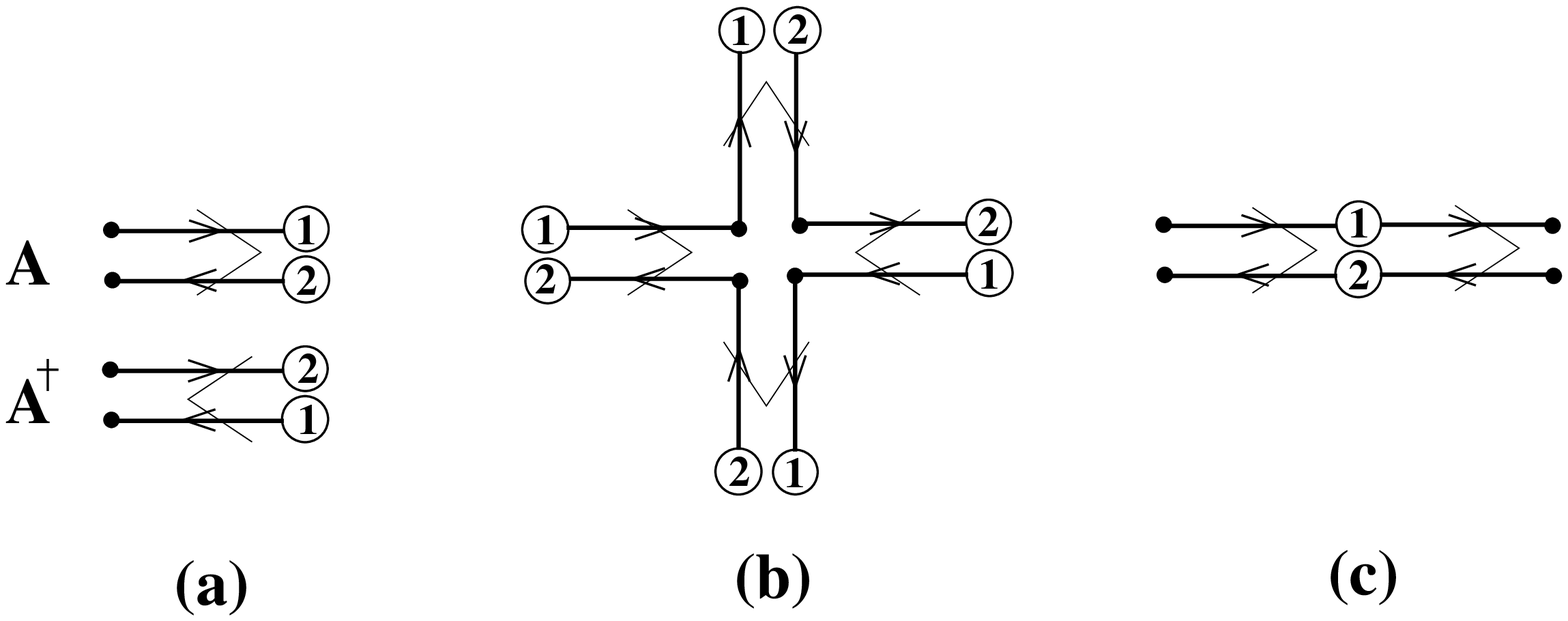}{10.cm}
\figlabel\combrect

\fig{A typical planar graph obtained from rectangular matrix integration.
The edges of the graph are made of two oriented lines, each carrying
a matrix index running respectively from $1$ to $N_1$ and $N_2$.
The faces of the graph are colored in white (resp. grey) according to the
index range $N_1$ (resp. $N_2$) of the oriented line
bordering it. With the above rules, face colors must alternate around each vertex.
Summing over running indices results in a weight $N_1$
per white face and $N_2$ per grey one, namely a weight $N_1^6N_2^7$ for the
present graph. }{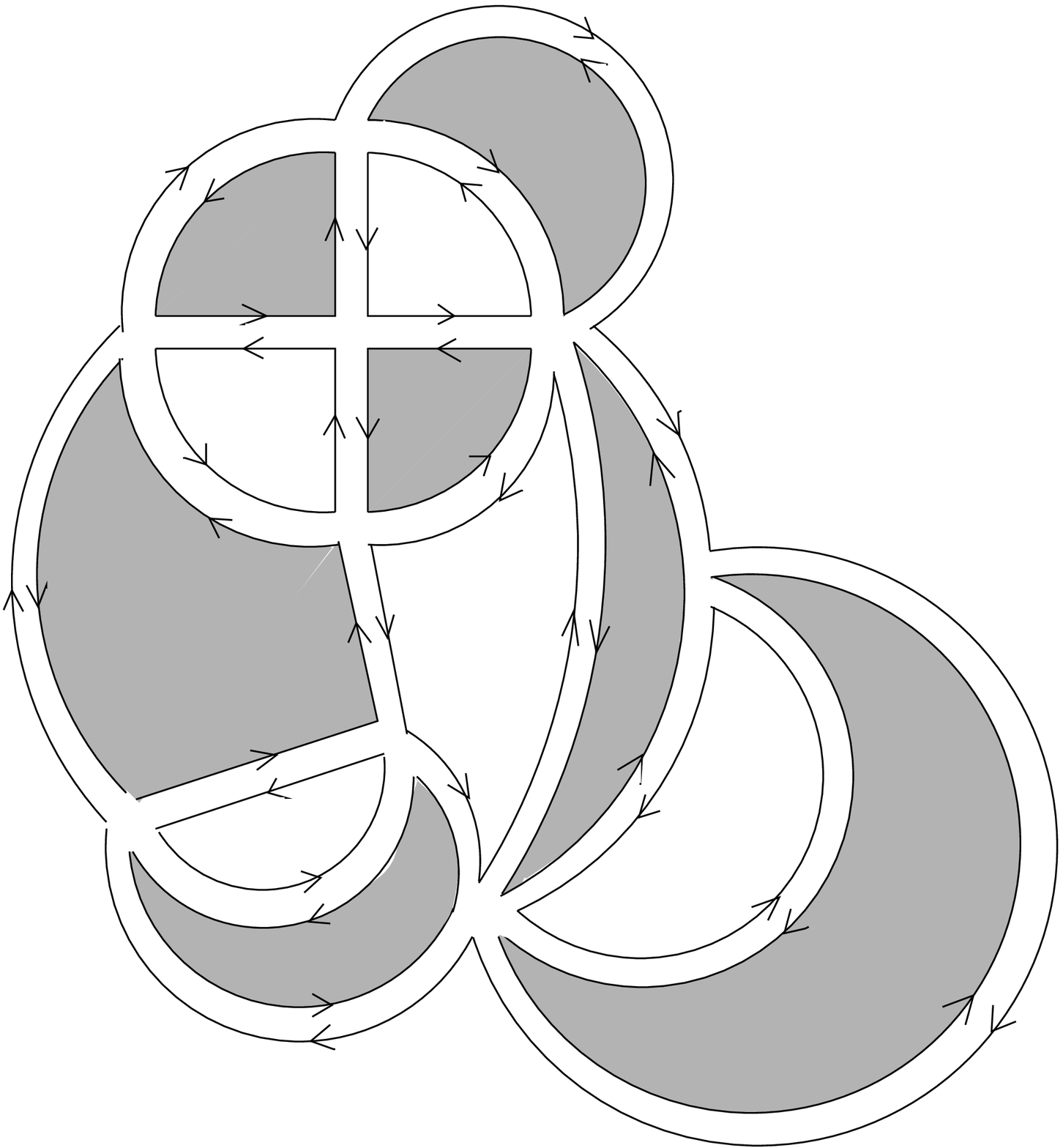}{8.cm}
\figlabel\bicofat

In this note we will concentrate 
on the combinatorial applications of rectangular matrix integrals.
For the purpose of illustration, we now consider an integral over complex 
rectangular matrices $A$ of size $N_1\times N_2$ of the form
\eqn\formrect{ Z={\int dA e^{-N{\rm Tr}({A A^\dagger\over t}-\sum_{k\geq 1}g_k {(AA^\dagger)^k
\over k} )} \over \int dA e^{-N{\rm Tr}({A A^\dagger\over t})} }}
where the measure of integration is flat over the complex elements: 
$dA=\prod_{i,j} dA_{ij} d {\bar A}_{ij}$.
Note that the integrand is unitary invariant like in \hermat: this is what makes the integral
easy to calculate, by changing variables to the eigenvalues of $AA^\dagger$.
We compute $Z$ again by expanding the $g_k$-terms in the exponential, and by
performing their term-by-term Gaussian integration. The latter is again obtained through
a similar graphical procedure to that of the Hermitian matrix case. 
We have represented in Fig. \combrect\ respectively
(a) the double half-edge for a matrix element $A_{ij}$, carrying the indices $i=1,2,...,N_1$
and $j=1,2,...,N_2$, where  the distinction between the two ranges of indices 
can be made by coloring the
two corresponding oriented lines say with colors $1$ and $2$; 
(b) the $2k$-valent vertex corresponding to a term ${\rm Tr}\big((AA^\dagger)^k\big)$ for $k=2$
and (c) the propagator $\langle A_{ij} A^\dagger_{kl}\rangle=\delta_{il}\delta_{jk}t/N$.  
Using again Wick's theorem,
each term in the expansion is integrated by connecting all half-edges into pairs so as to form
a closed graph like that of Fig. \bicofat. 
Note that the form of the propagator allows only to connect lines with indices
running over the same set. Summing over those results finally in a weight $N_i$ per corresponding
loop of color $i$. 
Equivalently, this amounts to coloring the faces of the graph with colors $1$ and $2$ (white and grey in
the graph of Fig. \bicofat),
and to weight them with $N_1$, $N_2$ respectively.  
Writing $N_i= Nx_i$ for convenience, and expanding ${\rm Log}\, Z$,
we get the generating function for arbitrary face-bicolored graphs with a weight
$t$ per edge, $x_i$ per face of color $i=1,2$, $g_k$ per $2k$-valent vertex, $k=1,2,...$,
an overall factor of  $N^{2-2h}$,  $h$ the genus of the graph and finally the usual inverse symmetry factor.

It is now clear how rectangular matrices will help solve colored-graph combinatorial problems. 
But before we go into this, we need a bit more experience with rectangular matrix integration,
the subject of next section.

\newsec{Rectangular Matrix Models}

In this section we recall a few known facts \AMP\ on complex rectangular
matrix integration, and apply them to the warm-up exercise of counting 
face-bicolored graphs with one vertex. 

\subsec{Reduction to eigenvalues}

As opposed the case of Hermitian (hence square) 
matrices, where the first step of integration consists in 
reducing the integral over the matrix to one over its eigenvalues,
this is no longer possible with a rectangular matrix $A$ as such.
However both $AA^\dagger$ and $A^\dagger A$ are Hermitian with non-negative
eigenvalues, and one may still reduce the rectangular matrix
integrals at hand to integrals over these non-negative eigenvalues.

More precisely, let us consider the general unitary invariant rectangular one-matrix
integral
\eqn\invint{ Z_N(V;N_1,N_2)= \int dA e^{-N{\rm Tr}\, V(AA^\dagger) } }
where the integral extends over the complex rectangular matrices $A$ 
of size $N_1\times N_2$, and we use the flat measure $dA=\prod_{i=1}^{N_1}\prod_{j=1}^{N_2}
dA_{i,j} d{\bar A}_{i,j}$. We have picked the normalization of the measure arbitrarily, 
as we will only be interested in
computing ratios of objects of the form \invint\ corresponding to different $V$'s.
Let us assume $N_1\leq N_2$. Diagonalizing the $N_1\times N_1$ Hermitian matrix $AA^\dagger$
leads to $AA^\dagger = P \Lambda^2 P^\dagger$, where $\Lambda={\rm diag}(\lambda_1,...,\lambda_{N_1})$,
a diagonal matrix with non-negative entries, and where 
$P\in U(N_1)$ is unitary. Repeating this for $A^\dagger A$ leads to
another diagonal matrix containing the $\lambda^2$'s and completed by $N_2-N_1$
zero eigenvalues. Upon rearranging the eigenvalues by permutation, we may write
$A^\dagger A = Q (\Lambda')^2 Q^\dagger$, where $\Lambda'={\rm diag}(\lambda_1,...,\lambda_{N_1},0,...,0)$
and $Q\in U(N_2)$. Upon picking positive square roots of the eigenvalues and modifying
$P$ and $Q$ accordingly, we finally
arrive at the ``rectangular diagonalization" formula 
\eqn\rectdiag{ A= P X Q^\dagger, \quad X=\pmatrix{
\lambda_1 & 0         & \cdots &   0           & 0      & \cdots & 0\cr
0         & \lambda_2 & \ddots & \vdots        & \vdots &        & \vdots\cr
\vdots    &           & \ddots & 0             & \vdots &        & \vdots  \cr 
0         & \cdots    &  0     & \lambda_{N_1} & 0      & \cdots & 0  \cr} } 
where $X$ has size $N_1\times N_2$.
It is then tempting to change the variables of integration in \invint\ to $A\to (P,Q,\Lambda)$,
and noting that the integrand only depends on $\Lambda$, to integrate out the angular
variables $P,Q$. This can indeed be done but we must be careful with the change of variables
itself, which is not bijective as such. Indeed, if $\Phi$ denotes an arbitrary diagonal
matrix of size $N_1\times N_1$ with diagonal entries of the form $e^{i\varphi_j}$, and $R$ an
arbitrary unitary matrix of size $(N_2-N_1)\times (N_2-N_1)$, we see that
the transformations 
\eqn\change{ P\to P \, \Phi  , \qquad Q \to Q \pmatrix{\Phi & 0 \cr 0 & R\cr} }  
leave $A$ unchanged. Therefore the right locus for the angular integration is
$(P,Q)\in \big(U(N_1)\times U(N_2)/U(N_2-N_1)\big)/U(1)^{N_1}$. 
The Jacobian of the transformation is then easily computed (see the appendix of Ref. \MOR\
for the case $N_1=N_2$, and Ref. \AMP\ for the straightforward generalization 
to $N_1\geq N_2$) and after 
integrating out the angular variables we get up to some proportionality constant
(the volume of the angular variables locus) 
\eqn\varchan{ Z_N(V;N_1,N_2) 
= 
\left(\prod_{i=1}^{N_1} \int_0^\infty d\lambda_i \lambda_i^{N_2-N_1} e^{-NV(\lambda_i) }\right)
\Delta(\Lambda)^2 }
where $\Delta(\Lambda)=\prod_{1\leq i<j\leq N_1} (\lambda_j-\lambda_i)$ is the Vandermonde
determinant of $\Lambda$. The two notable differences with the Hermitian case are that
(i) the $\lambda_i$'s are all non-negative and (ii)  the potential receives an extra
logarithmic piece from the Jacobian, namely $W(\lambda)=V(\lambda)-{N_2-N_1\over N}
{\rm Log}\, \lambda$,
and this is the only place where $N_2$ appears, while $N_1$ is the number of integration
variables $\lambda_i$. 
It is worth noting that the form of the potential $W$ obtained here is similar to that
of the so-called ``generalized Penner model" studied in Ref. \CHAR\ (with a parameter
$t=-(N_2-N_1)/N$) where however it is used to define (square) Hermitian matrix integrals.

\subsec{Methods for solving}

The next step uses the well known property of the Vandermonde determinant
that for any monic polynomials $p_m(\lambda)=\lambda^m+O(\lambda^{m-1})$ we have
$\Delta(\Lambda)=\det(\lambda_i^{j-1})=\det(p_{j-1}(\lambda_i))$. We may now
conveniently pick the $p_m$'s to be orthogonal \wrt to the one-dimensional measure  
$d\lambda w(\lambda)= d\lambda \lambda^{N_2-N_1} e^{-NV(\lambda)}$, namely
\eqn\ortopol{ (p_n,p_m)\equiv \int_0^\infty d \lambda  \lambda^{N_2-N_1} e^{-NV(\lambda)}
p_n(\lambda)p_m(\lambda)=h_n\delta_{n,m}}
for some normalization factors $h_n$,
and finally expanding the determinants one gets 
\eqn\partfunch{Z_N(V;N_1,N_2)=N_1! \prod_{j=0}^{N_1-1} h_j }
The partition function $Z_N(V;N_1,N_2)$ is then explicitly computed by deriving recursion 
relations between the $h_n$'s from their definition \ortopol. In the following sections, we will 
solve some explicit models for combinatorial purposes.

In this paper we will be focusing on combinatorial applications of the partition
function \invint\ for various choices of the potential $V$, by interpreting them as suggested above
as generating functions for colored graphs.
In particular, we will be interested in the
limit of planar (genus zero) graphs, obtained in the large $N$ limit as explained above. 
Introducing $x_i=N_i/N$, for $i=1,2$, the planar limit is 
obtained by sending $N\to \infty$, while $x_1,x_2,V$ remain fixed, and we will assume
that $x_1<x_2$.
By a slight abuse of notation, we denote by $Z_N(V;x_1,x_2)\equiv Z_N(V;N_1,N_2)$.
We whish to compute the planar free energy for the correctly normalized model, namely
with partition function \formrect\ expressed as $Z_N(V;x_1,x_2)/Z_N(V_0;x_1,x_2)$, 
where $V_0(x)=x/t$ is the pure Gaussian potential.
Defining the following planar free energies:
\eqn\planfren{\eqalign{ f&\equiv f(V;x_1,x_2)= 
\lim_{N\to \infty} {1\over N^2} \, {\rm Log}\, Z_N(V;x_1,x_2)\cr
f_0&\equiv f(V_0;x_1,x_2)\cr}}
the correct generating function for bi-colored planar graphs is simply $f-f_0$. 

In the large $N$ limit, $Z_N(V;x_1,x_2)$ may also be computed by a saddle-point method.
Indeed it reads $Z_N(V;x_1,x_2)=\int d\lambda_1 ...d\lambda_{N_1} 
e^{-N^2 S(\lambda_1,...,\lambda_{N_1})}$, where the action functional $S$ is defined as 
\eqn\actfunc{ S(\lambda_1,...,\lambda_{N_1})={1\over N}\sum_{i=1}^{N_1}
\bigg\{ V(\lambda_i) -(x_2-x_1) {\rm Log}\, \lambda_i\bigg\} -{1 \over N^2}\sum_{1\leq i\neq j\leq N_1}
{\rm Log}|\lambda_i-\lambda_j|}
The large $N$ behavior is then dominated by the solution to the saddle-point equation 
$\partial S/\partial \lambda_i=0$, namely
\eqn\sapo{ V'(\lambda_i) -{x_2-x_1\over \lambda_i} ={2\over N} \sum_{j\neq i}
{1\over \lambda_i-\lambda_j} }
Let us as usual introduce the resolvent
\eqn\reso{ \omega_N(z)= {1\over N} \sum_{i=1}^{N_1} {1\over z-\lambda_i}}
expressed in terms of the solution to \sapo. 
When expanded at large $z$, this function generates the 
quantities 
\eqn\coeffom{\omega_{N,m}\equiv {1\over N}\langle {\rm Tr}\big((AA^\dagger)^m\big)\rangle
={1\over N}{\int dA {\rm Tr}((AA^\dagger)^m) e^{-NV(AA^\dagger)}
\over \int dA e^{-NV(AA^\dagger)} } }
through $\omega_N(z)=\sum_{m\geq 0} \omega_{N,m} z^{-m-1}$, with leading term
$\omega_{N,0}=N_1/N=x_1$.
Equivalently we may define the density of eigenvalues
$\rho_N(\lambda)={1\over N}
\sum_{i=1}^{N_1} \delta(\lambda-\lambda_i)$, a distribution related to 
the resolvent through
\eqn\relrhoom{ \rho_N(\lambda)={1\over 2i\pi}(\omega_N(\lambda+i0)-\omega_N(\lambda-i0))}
Multiplying both sides of \sapo\
with $1/(z-\lambda_i)$ and summing over $i$, we arrive at the following
quadratic differential equation for the resolvent:
\eqn\quagenv{{1\over N} {d\omega_N(z)\over dz}
+ \omega_N(z)^2 -W'(z) \omega_N(z) +P_N(z) =0 }
where $W(z)=V(z)-(x_2-x_1) {\rm Log}\, z$ is the total potential including the extra
logarithmic term from the Jacobian, and 
\eqn\defP{\eqalign{
P_N(z)&={1\over N} \sum_{i=1}^{N_1} {W'(z)-W'(\lambda_i)\over z-\lambda_i} \cr
&=-{x_2-x_1\over z}\omega_N(0)
+{1\over N} \sum_{i=1}^{N_1} {V'(z)-V'(\lambda_i)\over z-\lambda_i}\cr} }
When $N\to \infty$, the derivative term in \quagenv\ vanishes and we end up with a quadratic
equation for the resolvent $\omega(z)\equiv \lim_{N\to \infty} \omega_N(z)$, easily
solved as
\eqn\solgenres{ \omega(z)= {W'(z)\pm \sqrt{W'(z)^2 -4  P(z)} \over 2 } }
where the $\pm$ sign is selected by demanding that $\omega(z)\sim x_1/z$ at large $z$
(a property clearly inherited from the definition \reso). To further fix $\omega(z)$ completely,
we will make the standard ``one-cut" hypothesis, that this complex function only has one
cut in the complex plane, translating into the fact that the limiting density of eigenvalues
$\rho(\lambda)=\lim \rho_N(\lambda)$ has a compact support made of a single real interval.

To finally recover the planar free energy \planfren, let us  
note that $f=-\lim_{N\to \infty} S(\lambda_1,...,\lambda_{N_1})$, evaluated on the solution of the
saddle-point equation \sapo, we  then easily compute
\eqn\frencalc{ t{df\over dt}=-t{\partial S\over \partial t}
-\sum_{i=1}^{N_1} {\partial S\over \partial \lambda_i}\,  
t{d \lambda_i\over dt}={1\over t}\lim{1\over N} 
\sum_ {i=1}^{N_1}\lambda_i={\omega_1\over t}}
where we denote by $\omega_m=\lim_{N\to \infty} \omega_{N,m}$, $\omega_{N,m}$ as in \coeffom. 
So finally the corresponding generating function for bi-colored graphs with a marked edge
reads
\eqn\genE{ E(t)=  t{df\over dt}-t{df_0\over dt}={1\over t}(\omega_1(V)-\omega_1(V_0))}
where the mention of $V$ and $V_0$ simply refer to the potential one must use to compute
the corresponding coefficient of $z^{-2}$ in the resolvent's expansion at large $z$.

In the following sections, we will complete the calculations outlined here in the case of some
specific potentials $V$. Let us for starters deal with the Gaussian case $V=V_0$.

\subsec{A first combinatorial application: bicolored trees and the Gaussian model}

The simplest rectangular matrix model is the Gaussian one, say with potential
\eqn\potgo{ V_0(AA^\dagger)=AA^\dagger/t}
The corresponding orthogonal polynomials $p_n$
(satisfying \ortopol\ with $V=V_0$) are expressed in terms of Laguerre polynomials
and lead to a trivial result for $Z_N(V_0;x_1,x_2)\propto t^{N_1N_2}$.
It is however instructive to derive first the large $N$ (planar) limit of the model, 
to see in particular how the Wigner's semi-circle limiting distribution of 
eigenvalues for the Hermitian case is modified in the complex rectangular case. 
Moreover, the corresponding resolvent is the generating
function of face-bicolored planar graphs with one vertex, in bijection with
bicolored trees.

The quadratic equation \quagenv\ for the resolvent reads in the large $N$ limit
\eqn\sapogo{ \omega^2-\big({1\over t}-{x_2-x_1\over z}\big) \omega+P(z)=0 }
Moreover we have 
\eqn\pexpli{ P_N(z) ={x_2-x_1\over z} {1\over N} \sum_{i=1}^{N_1}{1\over \lambda_i}= 
-{x_2-x_1\over z}\omega_N(0) \to P(z)=-{x_2-x_1\over z}\omega(0) }
so that finally
\eqn\gogo{ \omega(z)= {1\over 2 }\left({1\over t} -{x_2-x_1\over z} -
\sqrt{\left({1\over t} -{x_2-x_1\over z}\right)^2+4 {x_2-x_1\over z}\omega(0)} \right) }
The negative branch of the square root has been selected to ensure that $\omega(z) \sim x_1/z$ 
for large $z$, which further fixes $\omega(0)=-1/(t (x_2-x_1))$ and finally
\eqn\fingo{ \omega(z)= {1\over 2 t z}\left(z -t(x_2-x_1) -
\sqrt{z^2 -2t(x_1+x_2) z+t^2(x_2-x_1)^2}\right) } 
The corresponding limiting density of eigenvalues reads
$\rho(\lambda)=(\omega(\lambda+i0)-\omega(\lambda-i0))/(2i\pi)$, so that
\eqn\rhogo{ \rho(\lambda)= {1\over 2 \pi t \lambda} 
\sqrt{2t(x_1+x_2) \lambda-\lambda^2-t^2(x_2-x_1)^2} }
This distribution has 
a compact support $\lambda\in [(\sqrt{tx_2}-\sqrt{tx_1})^2,(\sqrt{tx_2}+\sqrt{tx_1})^2]$, 
away from the origin as $x_2>x_1$.
The corresponding planar free energy $f_0$ obeys eqn. \frencalc, with the result
\eqn\frezero{ t {df_0\over dt}= {\omega_1\over t}=x_1x_2 }
Note that in the limit $x_2\to x_1\to 1$, upon performing the change of variables $\lambda=y^2$
the measure $\rho(\lambda)d\lambda$ reduces to the Wigner-semicircular distribution,
with support $[-2\sqrt{t},2\sqrt{t}]$.

The large $N$
resolvent $\omega(z)=\sum_{k\geq 0} \omega_k z^{-k-1}$ may be interpreted as the 
generating function for some bicolored graphs
as follows. 
Indeed, in the planar limit, the coefficients $\omega_k=\lim_{N\to \infty} 
\langle {\rm Tr}(M^k)\rangle/N$ 
are obtained by drawing all planar, face-bicolored
graphs, with one $2k$-valent vertex, and with a weight 
$x_i$ per face of color $i$, and $t$ per edge. 
Expanding $\omega(z)$ we recover the result of \DEG:
\eqn\resdeg{ \omega_k
=t^k\sum_{m=0}^{k-1} {1\over k} {k\choose m}{k-1\choose m} x_1^{k-m} x_2^{m+1}}
where these graphs were identified as bicolored arch configurations, equivalent
by duality to bicolored rooted trees.

\subsec{All-genus Gaussian model: face-bicolored graphs with one vertex}

We now wish to compute the all genus expansion  
\eqn\qties{ \theta_k= {1\over N}\langle {\rm Tr} (AA^\dagger)^k \rangle \equiv
{1\over N} {\int dA {\rm Tr} (AA^\dagger)^k e^{-N {\rm Tr} AA^\dagger/t} \over
\int dA e^{-N {\rm Tr} AA^\dagger/t} }=
\sum_{h\geq 0} N^{-2h} \theta_{k,h} }
where $\theta_{k,h}$ is the number of genus $h$ face-bicolored graphs with one 
$2k$-valent vertex, and with a weight $t$ per edge and $x_i$ per face of color $i=1,2$.
In the previous section, we have computed the large $N$ limit
$\omega_k=\theta_{k,0}$ \resdeg.
To go beyond this requires to go deeper into the
orthogonal polynomial solution of the model. The orthogonal polynomials $p_n$
for the Gaussian case are nothing but a rescaled version of
the generalized Laguerre polynomials $L_n^{\alpha}(x)$, orthonormal \wrt the measure
$dx x^\alpha e^{-x}$ on $\IR_+$. More precisely, we have the generating function \BAT: 
\eqn\genlag{G(x,z)\equiv e^{{N\over t} {xz \over 1+z}} =\sum_{n=0}^\infty {p_n(x)\over n!} 
\left({N\over t}\right)^n z^n }
with $p_n$ as in \ortopol, with $V=V_0$: the orthogonality property is readily obtained
by computing
\eqn\geneint{\eqalign{ \int_0^\infty dx x^{N_2-N_1}&e^{-Nx/t} G(x,z)G(x,w)=
\sum_{n,m\geq 0} {z^n w^m\over n! m!} \left({N\over t}\right)^{n+m}(p_n,p_m)\cr
&= (N_2-N_1)! \left({t/N\over 1-zw}\right)^{N_2-N_1+1} \cr} }
The fact that the \rhs of \geneint\ is a function of $zw$ implies the
orthogonality, and this immediately yields the normalization factors $h_n$
upon expanding the \rhs of \geneint\ with the result
\eqn\reshn{ h_n= \left({t\over N}\right)^{N_2-N_1+2n+1} \, n! (N_2-N_1+n)! }
which finally yields
\eqn\gaures{ Z_N(V_0;x_1,x_2)=N_1! \prod_{n=0}^{N_1-1} h_n = \left({t\over N}\right)^{N^2 x_1x_2}
N_1! \prod_{n=0}^{N_1-1} n! (N(x_2-x_1)+n)!} 
where we recover the expected $t^{N_1N_2}$ behavior.

To compute $\theta_k$, we use the following generating function
\eqn\gentet{ H(u)\equiv \sum_{k\geq 0} {(Nz/t)^k\over k!}\theta_k={1\over N} 
\langle {\rm Tr}( \, e^{Nu AA^\dagger/t}) \rangle} 
Reducing to eigenvalues,
it is a simple exercise to show that for any function $f$, the average 
$\langle f \rangle\equiv \langle \sum_{i=1}^{N_1}f(\lambda_i)\rangle$ is 
expressed in terms of the orthogonal polynomials as
\eqn\avort{ \langle f \rangle = 
\sum_{n=0}^{N_1-1} {(f(x)p_n,p_n)\over (p_n,p_n)} }
Applying this to $f(x)=e^{Nux/t}$, we are left with the computation
of the modified norms ${\tilde h}_n=(e^{Nux/t}p_n,p_n)$, while $h_n=(p_n,p_n)$
are given by \reshn. Using again the generating function $G(x,z)$ \genlag,
we write
\eqn\wewrite{\eqalign{ \int dx x^{N_2-N_1} &e^{N(u-1)x/t} G(x,z)G(x,w) =
\sum_{n,m\geq 0} {z^n w^m \over n! m!} (e^{Nux/t}p_n,p_m) \left({N\over t}\right)^{n+m}\cr
&=(N_2-N_1)! \left({t/N \over 1-z w-u(1+z)(1+w)}\right)^{N_2-N_1+1} \cr}}
Picking the diagonal terms in the $z,w$ expansion, we finally get a closed formula
\eqn\clotet{ {\tilde h}_n= (n!)^2\left({t\over N}\right)^{N_2-N_1+2n}  
\sum_{m\geq 0} {u^m\over m!} \sum_{r=0}^{min(n,m)}
{m\choose r}^2 {(N_2-N_1+m+n-r)!\over (n-r)!} }
which together with \reshn\ and \avort\ allows to compute the
generating function $H(u)$ of \gentet\ as:
\eqn\tetclo{ H(u)= {1\over N}\sum_{n=0}^{N_1-1} \sum_{0\leq r\leq m \leq r+n} 
u^m {m\choose r}{n\choose m-r}{N_2-N_1+n+r\choose r} }
and finally picking the coefficient of $u^k$, we get
\eqn\thetfin{ \theta_k= {k! t^k \over N^{k+1}} \sum_{n=1}^{N_1}\sum_{r=max(0,k+n-N_1)}^k
{k\choose r}{N_1-n\choose k-r}{N_2-n+r\choose r}}
A first remark is in order. If we suppress the coloring, by taking $N_1=N_2=N$, we
get a different result from that of Ref. \ITZIZUB\ obtained in the context of modular 
group combinatorics, enumerating the all-genus graphs with one $2k$-valent vertex:
indeed, arbitrary graphs with one $2k$-valent vertex are not necessarily bicolorable
as soon as their genus $h\geq 1$.

\fig{The only genus one face-bicolored graph with one
hexavalent vertex has one face of each color, and corresponds
to the term $x_1x_2/N^2$ in $\theta_3$.}{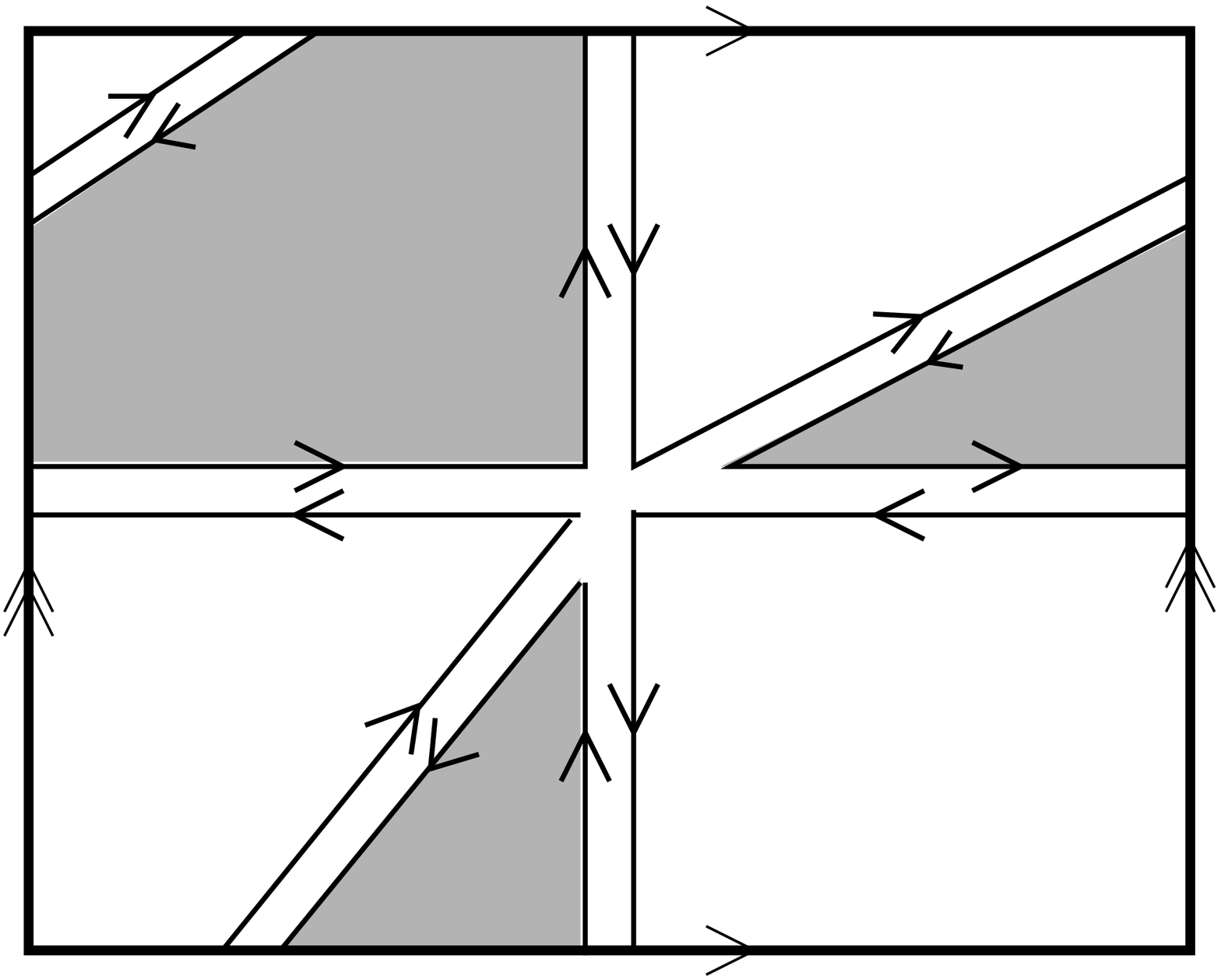}{6.cm}
\figlabel\gift

The first few numbers $\theta_k$ read
\eqn\firfew{\eqalign{
\theta_1&=x_1x_2 \cr
\theta_2&=x_1x_2(x_1+x_2)\cr
\theta_3&=x_1x_2(x_1^2+3 x_1x_2+x_2^2+{1\over N^2})\cr
\theta_4&=x_1x_2(x_1+x_2)(x_1^2+5x_1x_2+x_2^2+{5 \over N^2})\cr
\theta_5&=x_1x_2(x_1^4+10x_1^3x_2+20x_1^2x_2^2+10x_1x_2^3+x_2^4
+{5\over N^2}(3x_1^2+8x_1x_2+3x_2^2)+{8\over N^4})\cr}}
As an illustration, we have represented the first genus one contribution
appearing in $\theta_3$ at order $1/N^2$ in Fig. \gift.

\newsec{The one-matrix model: vertex-bicolored graphs}

Let us now consider the case of a general potential
\eqn\genepot{ V(AA^\dagger)= {AA^\dagger\over t} -\sum_{k=1}^M {g_k \over k} (AA^\dagger)^k }
The combinatorial interpretation of $Z_N(V;x_1,x_2)/Z_N(V_0;x_1,x_2)$ is clear: 
it generates arbitrary 
face-bicolored graphs with weights $g_k$ per $2k$-valent vertex, $k=1,2,...,M$,
$t$ per edge and $x_i$ per face of color $i=1,2$. Taking a finite upper bound $M$ is just an artifact
to make calculations easier, but our result will extend to any formal power series involving
all $g_k$, $k\geq 1$.
The section is organized as follows. We first solve the problem by use of orthogonal
polynomial techniques. The result is a set of recursion relations determining the free energy
of the model, that generates connected face-bicolored graphs of arbitrary genus.
In the subsequent section, we concentrate on the planar (large $N$) limit and derive closed
expressions for the corresponding generating functions.

\subsec{Orthogonal polynomial solution}

We wish to compute the all-genus free energy
\eqn\allgen{\eqalign{ F_N(V;x_1,x_2)&={1\over N^2} \, {\rm Log}\left({Z_N(V;x_1,x_2)
\over Z_N(V_0;x_1,x_2)}\right)\cr
&=\sum_{h\geq 0} N^{-2h} f_h(V;x_1,x_2) \cr}}
where $f_h$ is the generating function for connected face-bicolored graphs of genus $h$
with a weight $t$ per edge, $g_k$ per $2k$-valent vertex and $x_i$ per face of color $i=1,2$.
The relation \partfunch\ allows to compute $F_N$ completely in terms of the normalizations
$h_n(V),h_n(V_0)$ of the corresponding orthogonal polynomials \ortopol.
To determine relations between the $h$'s, we compute
\eqn\compute{\eqalign{
{\partial \over \partial t} (p_n,p_n) &= 2({\partial p_n\over \partial t},p_n)
+{N\over t^2} (xp_n,p_n)\cr
{\partial \over \partial t} (p_n,p_{n-1}) &=({\partial p_n\over \partial t},p_{n-1})
+{N\over t^2}(xp_n,p_{n-1})=0\cr}}
The operator of multiplication by $x$ acting on the $p_n$'s is self-adjoint and
therefore reads
\eqn\multx{ xp_n(x)=p_{n+1}(x)+ s_n p_n(x)+r_n p_{n-1}(x), \qquad r_n={h_n\over h_{n-1}}}
Moreover, if we write $p_n(x)=x^n-\lambda_n x^{n-1}+O(x^{n-2})$ and substitute it into
\multx, we easily identify $\lambda_n=\sum_{j=0}^{n-1} s_j$. Finally, by using the orthogonality
of the $p$'s, we see that $({\partial p_n\over \partial t},p_n)=0$ and that
$({\partial p_n\over \partial t},p_{n-1})=-\lambda_n h_{n-1}$. This allows to rewrite
\compute\ as
\eqn\putecom{\eqalign{ {t^2\over N^2}{\partial \over \partial t}{\rm Log}\, h_n &=   
s_n \cr
{t^2\over N^2}{\partial \over \partial t}\lambda_n&=\sum_{j=0}^{n-1} 
{t^2\over N^2}{\partial \over \partial t} s_j= r_n \cr}}
Introducing the ``partial" free energy $F_n=(1/N^2)\sum_{0\leq j\leq n-1}{\rm Log}\, h_j$,
we may write $r_n=(t^2\partial_t)^2F_n/N^2$ which finally leads to
a closed equation for $F_n$, by noting that $F_{n+1}+F_{n-1}-2F_n=({\rm Log}\,r_n)/N^2$:
\eqn\closedF{ e^{N^2(F_{n+1}+F_{n-1}-2F_n)} = {1\over N^2}
\left(t^2{\partial\over \partial t}\right)^2 F_n }
Together with the initial conditions $F_0=0$ and 
$F_1=(1/N^2){\rm Log}\,h_0$, this determines $F_n$ completely
as a power series of $t$, and allows to compute the desired all-genus free energy 
$F_N(V;x_1,x_2)=\big(F_{n}(V)-F_{n}(V_0)\big)\vert_{n=N_1}$, by using eqn. \closedF\
in both cases of $V$ and $V_0$.
More precisely, let us introduce the subtracted partial free
energy $f_n\equiv F_n(V)-F_n(V_0)$. Using the last equation of \putecom\
for both $V$ and $V_0$, we may write
\eqn\defrho{\eqalign{ 
\rho_n\equiv {r_n(V)\over r_n(V_0)}&={(t^2\partial_t)^2 F_n(V) \over 
(t^2\partial_t)^2 F_n(V_0)}=1+{(t^2\partial_t)^2 f_n \over 
(t^2\partial_t)^2 F_n(V_0)}\cr
&=e^{N^2(f_{n+1}+f_{n-1}-2f_n)}\cr} }  
We now use eqn. \reshn\ to explicitly compute the partial Gaussian free energy
$F_n(V_0)=(1/N^2)\sum_{i=0}^{n-1}{\rm Log}\,h_i(V_0)$ as
\eqn\compufo{ F_n(V_0)
= (1/N^2)\sum_{i=0}^{n-1}{\rm Log}\left((t/N)^{N_2-N_1+2i+1}i! (N_2-N_1+i)!\right)} 
and finally the denominator in \defrho\ reads
\eqn\denorho{ (t^2 \partial_t)^2 F_n(V_0)= {n(N_2-N_1+n)\over N^2} t^2 }
which leads finally to the following recursion relation for $f_n$:
\eqn\recufn{ f_{n+1}+f_{n-1}-2f_n=
{1\over N^2} {\rm Log}\left(1+{N^2\partial_t t^2\partial_t f_n\over n(N_2-N_1+n)} \right) }
to be supplemented by the following initial conditions
\eqn\initconfn{ \eqalign{ f_0&=0 \cr
f_1&={1\over N^2}{\rm Log}\left({\int_0^\infty dx x^{N_2-N_1} e^{-Nx/t} 
e^{N\sum_i g_i {x^i \over i}}
\over \int_0^\infty dx x^{N_2-N_1} e^{-Nx/t}}\right) \cr}}
Note that the details of the potential, in particular the dependence on
the $g_i$'s is entirely contained in the initial condition \initconfn, which is 
supposed to be formally expanded as a power series of the $g_i$'s. 

To solve \recufn\-\initconfn, let us proceed order by order in $t$. Indeed writing
\eqn\writfnt{ f_n ={1\over N^2}\sum_{k=1}^\infty f_{n,k} t^k}
then eqn. \recufn\ implies the following recursion relation, upon expanding the logarithm:
\eqn\folrec{ f_{n+1,k}+f_{n-1,k}-2 f_{n,k}= \sum_{p\geq 1} {(-1)^p \over p\, n^p(N_2-N_1+n)^p} 
\sum_{k_i\geq 1,i=1,2,...,p \atop \Sigma_i k_i=k} \prod_{i=1}^p k_i(k_1+1) f_{n,k_i}   }
to be fed with the initial conditions \initconfn\ that
$f_{0,k}=0$ and that $f_{1,k}$ is identified with the coefficient of $t^k$ in the
expansion of $N^2 f_1$ of \initconfn. Moreover, it is easy to prove by induction
that all $f_{n,k}$ are polynomials of $n$ of degree $k+1$, and always have a factor
$n(N_2-N_1+n)$. 

Explicitly, for the first few values of $k$, we find
\eqn\fircond{\eqalign{N^2f_{n,1}&= n(N_2-N_1+n) g_1  \cr
N^2f_{n,2}&={1\over 2}n(N_2-N_1+n)\big(g_1^2+{N_2-N_1+2n\over N} g_2 \big) \cr
N^2f_{n,3}&={1\over 3}n(N_2-N_1+n)\big(g_1^3+{3\over N}(N_2-N_1+2n)g_1g_2 \cr
&+{1\over N^2}((N_2-N_1)^2+5n(N_2-N_1)+5n^2+1) g_3\big)\cr
N^2f_{n,4}&={1\over 4}n(N_2-N_1+n)\big(g_1^4+{6\over N}(N_2-N_1+2n)g_1^2g_2\cr
&+ {4\over N^2}((N_2-N_1)^2+5n(N_2-N_1)+5n^2+1) g_1g_3\cr 
&+ {1\over N^2}((2(N_2-N_1)^2+9n(N_2-N_1)+9n^2+1) g_2^2\cr
&+ {1\over N^3}(N_2-N_1+2n)((N_2-N_1)^2+7n (N_2-N_1)+7n^2+5) g_4\big)\cr
N^2f_{n,5}&={1\over 5}n(N_2-N_1+n)\big(g_1^5+{10\over N}(N_2-N_1+2n)g_1^3g_2\cr
&+{10\over N^2}((N_2-N_1)^2+5n(N_2-N_1)+5n^2+1) g_1^2g_3\cr 
&+{5\over N^3}(N_2-N_1+2n)((N_2-N_1)^2+7n (N_2-N_1)+7n^2+5) g_1g_4 \cr
&+{5\over N^2}((2(N_2-N_1)^2+9n(N_2-N_1)+9n^2+1) g_1 g_2^2\cr 
&+ {5\over N^3}(N_2-N_1+2n)((N_2-N_1)^2+6n (N_2-N_1)+6n^2+3) g_2g_3 \cr
&+{1\over N^4}((N_2-N_1)^4+14 n(N_2-N_1)^3+(56n^2+15)(N_2-N_1)^2\cr
&+14n(6n^2+5)(N_2-N_1)+2(21n^4+35n^2+4)) g_5 \big)\cr}}

\fig{Two genus one graphs corresponding to terms of order $1/N^2$ in $F_{N,4}$,
with respective weights $g_2^2 x_1 x_2$ (2 four-valent vertices,
one face of each color) and $g_4 x_1x_2^2$ (one eight-valent vertex,
one white and two grey faces).}{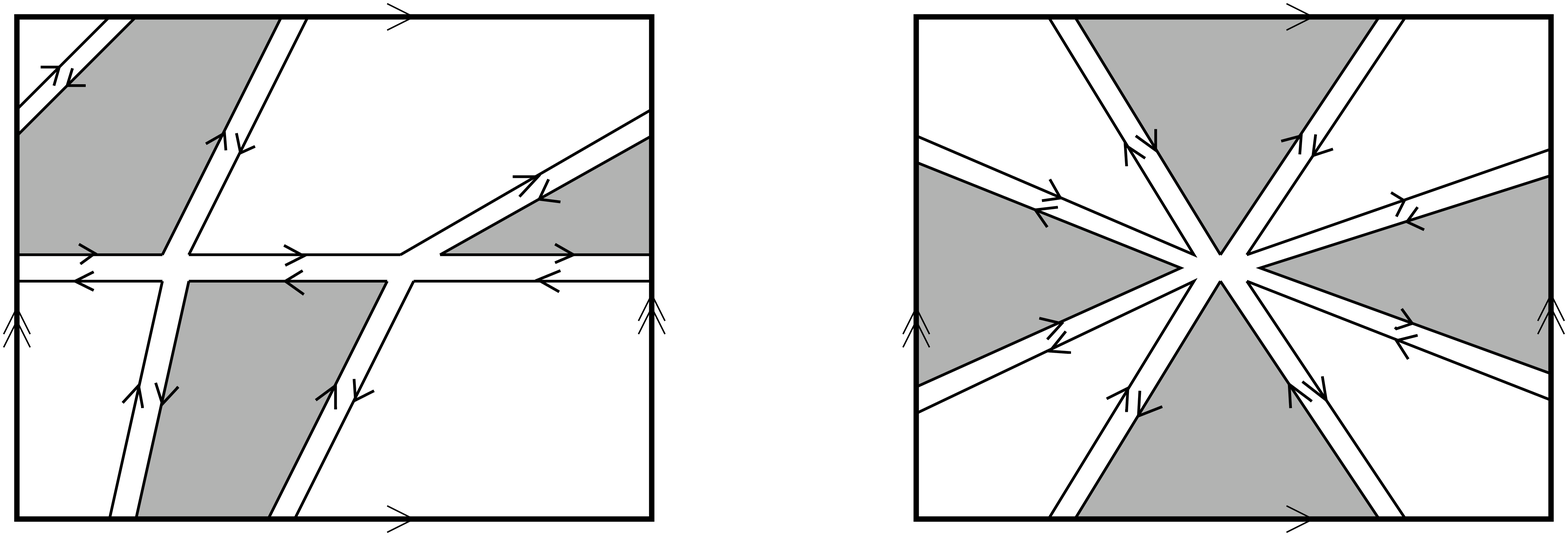}{10.cm}
\figlabel\newgift

To get the final result for the subtracted free energy 
$F_N(V;x_1,x_2)=\sum_{k\geq 1} F_{N,k} t^k$ of \allgen, 
we simply note that $F_{N,k}=f_{n,k}\vert_{n=N_1}$, hence we finally get
\eqn\firF{\eqalign{
F_{N,1}&=x_1x_2 g_1 \cr  
F_{N,2}&={1\over 2}x_1x_2\big(g_1^2+(x_1+x_2)g_2\big)\cr
F_{N,3}&={1\over 3}x_1x_2\big(g_1^3+3(x_1+x_2)g_1g_2
+(x_1^2+3x_1x_2+x_2^2+{1\over N^2})g_3\big)\cr
F_{N,4}&={1\over 4}x_1x_2\big(g_1^4+6(x_1+x_2)g_1^2g_2
+4(x_1^2+3x_1x_2+x_2^2+{1\over N^2})g_1 g_3 \cr
&+(2x_1^2+5x_1x_2+2x_2^2+{1\over N^2})g_2^2
+(x_1+x_2)(x_1^2+5x_1x_2+x_2^2+{5\over N^2})g_4\cr
F_{N,5}&={1\over 5}x_1x_2\big(g_1^5+10(x_1+x_2)g_1^3g_2
+10(x_1^2+3x_1x_2+x_2^2+{1\over N^2})g_1^2 g_3\cr
&+5(x_1+x_2)(x_1^2+5x_1x_2+x_2^2+{5\over N^2})g_1g_4
+5(2x_1^2+5x_1x_2+2x_2^2+{1\over N^2})g_1g_2^2\cr
&+5(x_1+x_2)(x_1^2+4x_1x_2+x_2^2+{3\over N^2})g_2g_3
+(x_1^4+10x_1^3x_2\cr
&+20x_1^2x_2^2+10x_1x_2^3+x_2^4+{5\over N^2}(
3x_1^2+8x_1x_2+3x_2^2)+{8\over N^4})g_5 \big)\cr}}
The quantities $F_{N,k}$ are the generating functions for arbitrary face-bicolored 
graphs with $k$ edges, and with a weight $N^{-2h}$, $h$ the genus, a weight $x_i$ per face of color $i$,
and a weight $g_i$ per $2i$-valent vertex, and also an inverse symmetry factor. 
We have displayed in Fig. \newgift\ for illustration some of the genus one graphs
contributing to $F_{N,4}$.
When $N\to \infty$, we may identify easily several coefficients in $F_{N,k}$.
In particular, the coefficient of $g_1^k$ simply counts graphs made of one
loop with $k$ marked points, i.e. with a symmetry factor $1/k$, and therefore two
faces, one of each color.
Moreover, the coefficient of $g_k$ in $F_{N,k}$ 
(all-genus graphs with one $2k$-valent vertex) is easily identified
with $\theta_k$ of \firfew.  

In the following section, we use an alternative method
to directly get the planar limit of the free energy in a more compact
form.

\subsec{Planar limit}

Repeating the computation of Sect.3.2, we finally arrive at the planar limit of the 
resolvent in the form
\eqn\forplares{ \omega(z)={zV'(z)-x_2+x_1 +\sqrt{(zV'(z)-x_2+x_1)^2-4 z^2P(z)}\over 2 z }}
where 
$zV'(z)-x_2+x_1$ is a polynomial of degree $M$ in $z$, and $zP(z)$ a polynomial of degree $M-1$.
Applying the ``one-cut" hypothesis, let us assume that the polynomial under
the square root has $M-1$ double zeros, and two single ones, say $b>a>0$, so that the density 
of eigenvalues has the compact support $[a,b]$. This amounts to writing
\eqn\amw{ \omega(z)={zV'(z)-x_2+x_1 + Q(z) \sqrt{(z-a)(z-b)} \over 2z } }
where $Q$ is a polynomial of degree $M-1$. Imposing that $\omega(z)\sim x_1/z$
at large $z$, we may solve for $Q$:
\eqn\solq{ Q(z)=-\left[ {zV'(z)-x_2+x_1 \over \sqrt{(z-a)(z-b)}} \right]_+ }
where the bracket must be expanded at large $z$, and the subscript $+$ simply means that 
only the polynomial part of this expansion must be retained. 
Note that, with these definitions, the term $-x_2+x_1$ in \solq\ may be dropped
as it only participates to terms of order $1/z$ and higher.
By analogy with the Gaussian result
\fingo, let us introduce the following parametrization for $a$ and $b$ 
\eqn\paramab{ a=(\sqrt{U_2}-\sqrt{U_1})^2 \ , \qquad b=(\sqrt{U_2}+\sqrt{U_1})^2 }
so that $a+b=2(U_1+U_2)$ and $ab=(U_2-U_1)^2$. Introducing the polynomial coefficients
$\psi_k(U_1,U_2)$ through the following large $z$ expansion 
\eqn\defpsi{\eqalign{ {1\over \sqrt{1-{2(U_1+U_2)\over z}+{(U_2-U_1)^2\over z^2}}}&=
\sum_{k\geq 0} {1\over z^k} \psi_k(U_1,U_2) \cr
\psi_k(U_1,U_2)&= \sum_{m=0}^k {k\choose m}^2 U_1^{k-m}U_2^m \cr}}
and substituting $V'(z)=1/t -\sum_{1\leq k\leq M} g_k z^{k-1}$ into \solq, we get
\eqn\qsolu{\eqalign{ Q(z)&=-\left[ {V'(z) \over \sqrt{1-{a+b\over z}+{ab\over z^2}}} 
\right]_+\cr
&= -{1\over t} +\sum_{m=0}^{M-1}
z^m \sum_{k=0}^{M-1-m} g_{m+k+1} \psi_k(U_1,U_2) \cr}}
To further fix $a$ and $b$, we must moreover write that $\omega(z)\sim x_1/z$ at large
$z$. Introducing the large $z$ series $R(z)=-V'(z)/\sqrt{1-{a+b\over z}+{ab\over z^2}}$,
we may rewrite 
\eqn\rewom{\omega(z)=(x_1-x_2+(Q(z)-R(z))\sqrt{(z-a)(z-b)})/(2 z ) }
so that the above condition reads
$x_1-x_2-R(z)\vert_{-1} =2x_1$ (the subscript simply means that we must pick the coefficient 
of the corresponding power of $z$ in the Laurent series for $R$), and eventually
\eqn\eventu{ x_1+x_2 ={1\over t} \psi_1(U_1,U_2) -\sum_{k=1}^M g_k \psi_{k}(U_1,U_2) } 
We must finally ensure that the resolvent $\omega$ is well-defined at $z=0$, which amounts to
imposing that the numerator of $\omega(z)$ in \amw\ vanishes\foot{We must pay attention here to
the fact that \amw\ defines $\omega$ correctly as a power series of $z$ at infinity. 
Going to the limit
$z\to 0$ implies an overall change of sign of the square root term.} 
at $z=0$, i.e. that
$-\sqrt{ab}\, Q(0)=x_2-x_1$, namely, using $\sqrt{ab}=U_2-U_1$:
\eqn\vanishres{ x_2-x_1=(U_2-U_1)\big({1\over t}-\sum_{k=0}^{M-1} g_{k+1} \psi_k(U_1,U_2)\big)} 
The two equations \eventu\-\vanishres\ determine $U_1$ and $U_2$ (i.e. $a$ and $b$) completely
upon requiring that $U_1,U_2$ be formally expanded in powers of $t$ around the origin, 
starting with $U_i=t x_i +O(t^2)$, $i=1,2$. These two equations may be recast into
\eqn\recasthem{\eqalign{
x_1&={U_1\over t} - \sum_{k=1}^M g_k \varphi_k(U_1,U_2) \cr 
x_2&={U_2\over t} - \sum_{k=1}^M g_k \varphi_k(U_2,U_1) \cr
\varphi_k(U_1,U_2)&=
\sum_{m=0}^{k-1} {k\choose m} {k-1\choose m} U_1^{k-m}U_2^{m} \cr}}
where the $\varphi_k$'s are generated by 
\eqn\genphik{{1\over 2}{1+{U_1-U_2\over z}\over \sqrt{1-2{U_1+U_2\over z}+{(U_1-U_2)^2\over z^2}}}
=\sum_{k\geq 0} {\varphi_k \over z^k} }

Let us now express the free energy of the model as in \frencalc\
$t {df \over dt}= {\omega_1\over t}$. 
We must use again the series $R(z)$ to express the $1/z^2$ 
coefficient of $\omega(z)$ in \rewom\ as 
$\omega_1= (-R(z)\vert_{-2} +{a+b\over 2} R(z)\vert_{-1})/2$.
Together with \frencalc, this finally yields
\eqn\finyiel{\eqalign{ t {df \over dt}&={U_1U_2\over t}
\big({1\over t}-\sum_{k=1}^{M} g_k \phi_k(U_1,U_2)\big)\cr
\phi_k(U_1,U_2)&= \sum_{m=0}^{k-1}
{k\choose m+1}{k \choose m} U_1^{k-m-1} U_2^{m} \cr}}
where the $\phi_k$'s are generated by
\eqn\genkphi{ {z\over 2 U_1U_2}\left({1-{U_1+U_2\over z}\over 
\sqrt{1-2{U_1+U_2\over z}+{(U_1-U_2)^2\over z^2}}} -1\right) =\sum_{k\geq 1} {\phi_k\over z^k} }  

We finally arrive at the complete generating function for rooted (edge-marked)
bicolored graphs with a weight $t$ per edge, $x_i$ per vertex of color $i=1,2$
and $g_k$ per $2k$-valent vertex, $k=1,2,...,M$:
\eqn\readfin{ E(t)=t {df \over dt} -t {df_0\over dt} ={U_1U_2\over t^2}
\big(1-t\sum_{k=1}^{M} g_k \phi_k(U_1,U_2)\big) -x_1 x_2 }
where $U_1,U_2$ are the unique solutions to \recasthem, such that $U_i=t x_i+O(t^2)$.
The result becomes even simpler if we compute the derivative
\eqn\derivEfin{ {d \over dt}(tE(t)) ={U_1U_2\over t^2}-x_1x_2  }
by using \recasthem\ to eliminate both $dU_1/dt$ and $dU_2/dt$.

A few remarks are in order. Eqns. \recasthem\ were first obtained 
in \SCH\ in a purely combinatorial way, by establishing bijections
between face-bicolored planar graphs and suitably colored trees, 
generated by the functions $U_1,U_2$ (denoted $tP,tQ$ there, while
$x_1,x_2$ are denoted $u,v$). 
Although neither expression \readfin\
nor \derivEfin\ appear there, we checked their compatibility with
the equation  $\partial E/\partial x_2=(U_1/t) -x_1$ obtained there
(actually expressed in terms of a function $B=E+x_1x_2$). 
Eqn. \derivEfin\ can be rephrased as a bijection between suitably marked 
face-bicolored graphs and pairs of such trees. 

As a non-trivial check of our calculations, let us
suppress the colors by taking $x_1=x_2=1$. We then have $U_1=U_2\equiv U$,
and $\varphi_k(U,U)= {2k-1\choose k} U^k$ from \genphik. This reduces eqn. \recasthem\ 
to eqn. (A.13) of appendix A upon redefining $M\to J$ and $g_i\to g_{2i}$. 
Noting that $\phi_k(U,U)={2k\choose k-1} U^{k-1}$ we also find
that \readfin\ reduces to eqn. (A.14)  of appendix A.
Finally, eqn. \derivEfin\ clearly reduces to eqn. (A.15) of appendix A.

As a final example, the generating function for rooted bicolored quadrangulations,
namely with $g_k=\delta_{k,2}$, reads
\eqn\rooquad{ \eqalign{ E(t)&= {U_1U_2\over t^2} (1-2t(U_1+U_2)) -x_1x_2\cr
x_1 &= U_1\big({1\over t}-(U_1+2 U_2)\big) \cr
x_2 &= U_2\big({1\over t}-(U_2+2 U_1)\big) \cr}}

\newsec{More matrices: vertex-tricolored triangulations}

\subsec{Rectangular multi-matrix model}

\fig{The pictorial representation for the triple Gaussian integration over $A_1,A_2,A_3$
with respective sizes $N_1\times N_2$, $N_2\times N_3$, $N_3\times N_1$. Each matrix
element is represented as a double half-edge (a), each oriented line
carrying one index (we have indicated the color $i=1,2,3$ of each index, corresponding
to its range $N_1,N_2,N_3$ respectively. Each half-edge carries an
additional overall orientation, to distinguish between $A_i$ and $A_i^\dagger$.
We have represented the two vertices Tr$(A_1A_2A_3)$ and 
Tr$(A_3^\dagger A_2^\dagger A_1^\dagger )$ in (b). The result of the integration is
to connect the half-edges into pairs by forming edges (c) along the lines of which indices
(and therefore index colors) are conserved.}{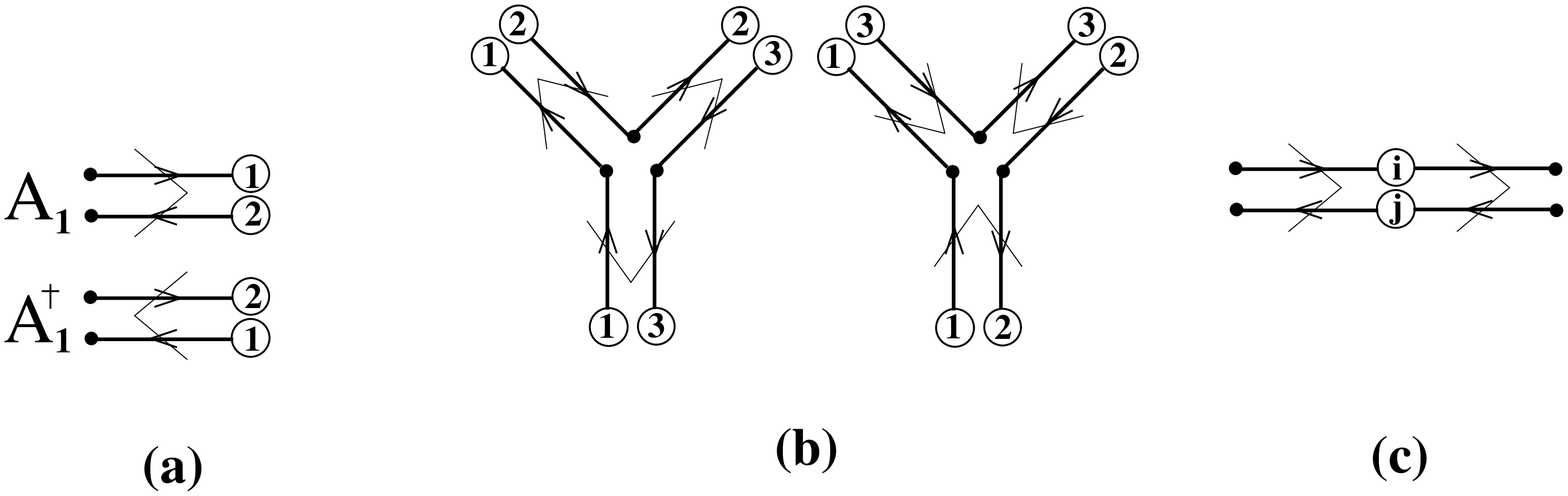}{12.cm}
\figlabel\trico

In \DEG, the generating function for vertex-tricolored random triangulations
with a weight $t$ per edge of color (12), and $x_i$ per vertex of color $i=1,2,3$
was obtained by use of a two-Hermitian matrix model. Here we present an
alternative solution, using the following integral over {\it three}
rectangular matrices $A_1,A_2,A_3$ respectively with sizes $N_1\times N_2$,
$N_2\times N_3$, $N_3\times N_1$:
\eqn\trivert{\eqalign{ Z_N(V;N_1,N_2,N_3)&=\int dA_1dA_2dA_3 e^{-N{\rm Tr}(V(A_1,A_2,A_3))}\cr
V(A_1,A_2,A_3)&= {1\over \tau}(A_1A_1^\dagger+A_2A_2^\dagger+A_3A_3^\dagger)
-A_1A_2A_3-A_3^\dagger A_2^\dagger A_1^\dagger \cr}}
The free energy of the model reads
\eqn\fretri{ F_N(V;N_1,N_2,N_3)-F_N(V_0;N_1,N_2,N_3)={1\over N^2}{\rm Log}\left({ Z_N(V;N_1,N_2,N_3)
\over  Z_N(V_0;N_1,N_2,N_3)}\right) }
where $V_0(A_1,A_2,A_3)= {1\over \tau}(A_1A_1^\dagger+A_2A_2^\dagger+A_3A_3^\dagger)$
is the Gaussian part of the potential.
The expansion of this free energy now involves summing over all connected trivalent fatgraphs,
with double-edges carrying indices ranging up to $N_i$, $i=1,2,3$ according to the corresponding
matrix indices. 
Indeed, expanding the non-Gaussian part of the exponential, we must again integrate term-by-term.
This is done diagrammatically using the elements of Fig. \trico:
(a) the matrix elements $(A_i)_{jk}$ are represented as double half-edges
as before, but now with three possible line colors $1,2,3$ corresponding to indices
running up to $N_1,N_2,N_3$ respectively; (b) the terms ${\rm Tr}(A_1A_2A_3)$ and
${\rm Tr}(A_3^\dagger A_2^\dagger A_1^\dagger)$ correspond to two trivalent vertices
around which colors cyclically alternate respectively
in clockwise and counterclockwise order; (c) finally the half-edges around the vertices
must be paired into edges through the propagators $\langle (A_m)_{ij} (A_m)_{kl}^\dagger \rangle=
\delta_{il}\delta_{jk}\tau/N$, in such a way that line colors are conserved. 

We therefore must draw all connected face-colored trivalent fatgraphs using colors $1,2,3$, such that
the three faces adjacent to each vertex have distinct colors. These graphs are nothing but the
duals of vertex-tricolored triangulations. 
Further setting $x_i=N_i/N$, $i=1,2,3$, we see that these graphs receive a weight $x_i$ per face of
color $i=1,2,3$, $\tau$ per edge, and an overall $N^{2-2h}$, $h$ the genus of the graph.   
The precise contact with the results of \DEG\ is made through setting $t=\tau^3$, as there are 
three times as many edges as those colored $(12)$ in the triangulations.
 
The integral \trivert\ may be simplified by explicitly integrating out say $A_3$.
Indeed, forming the product 
\eqn\forsq{
{1\over \tau}A_3A_3^\dagger- A_3A_1A_2- A_2^\dagger A_1^\dagger A_3^\dagger
={1\over \tau}(A_3-\tau A_2^\dagger A_1^\dagger)(A_3^\dagger-\tau A_1A_2)
-\tau A_2^\dagger A_1^\dagger A_1A_2}
and performing the Gaussian integration over $A_3'=A_3-\tau A_2^\dagger A_1^\dagger$, we are left
with a two-matrix integral with potential 
\eqn\leftwith{V(A_1,A_2)={1\over \tau}(A_1A_1^\dagger+A_2A_2^\dagger)
-\tau A_1^\dagger A_1A_2A_2^\dagger}
Now the integration over $A_2$ can also be explicitly performed, as a particular case of
the general formula
\eqn\genint{ {\int d A e^{-{\rm Tr}(AA^\dagger M)}\over \int d A e^{-{\rm Tr}(AA^\dagger)}}
={1\over \det(M)^{N_3} } }
where $A$ is rectangular complex of size $N_2\times N_3$ and $M$ non-singular complex of size
$N_2\times N_2$. We are finally left with a rectangular one-matrix integral over $A\equiv \tau A_1$
of size $N_1\times N_2$, and we may simply substitute in the logarithm of \fretri\
the ratio of partition functions with $Z_N(V)/Z_N(V_0)$ where
\eqn\wehav{\eqalign{ Z_N(V)&=\int dA e^{-NV(AA^\dagger)} \cr
V(AA^\dagger)&= {AA^\dagger \over t} +x_3\, {\rm Log}(I-AA^\dagger)\cr
V_0(AA^\dagger)&= {AA^\dagger \over t} \cr}} 
where we have set $t=\tau^3$.
So we finally see from simple direct matrix integrations 
that the counting of vertex-tricolored triangulations
reduces to that of bicolored Eulerian graphs, with a weight $x_3$ per vertex of any (even) degree, 
namely taking infinitely many $g_k$'s in \genepot, and taking them all equal to $x_3$. 
This relation was actually implicitly 
used in \DEG, where a two-Hermitian matrix model actually counted
the relevant Eulerian graphs. A more general relation was found in \BMS\ connecting the so-called
$k$-constellations to $k+1$-valent Eulerian graphs, and the present situation corresponds to $k=3$.

\subsec{Orthogonal polynomial solution}

At this point we just have to use the results of Sect.4.1, by simply setting $g_k=x_3$ for all
$k\geq 1$ in the various equations i.e. with $V(x)=x/t+x_3{\rm Log}(1-x)$. 
Some of them however must be handled with care.
Indeed, the main recursion relation \recufn\ remains the same, as it basically
is independent of the particular form of the potential $V$, but the initial condition
\initconfn\ is changed. It now reads
\eqn\newinit{\eqalign{f_0&=0\cr
f_1&={1\over N^2}{\rm Log}\left({\int_0^\infty dx x^{N_2-N_1}(1-x)^{-N_3}e^{-Nx/t}\over
\int_0^\infty dx x^{N_2-N_1} e^{-Nx/t}}\right)\cr
&={1\over N^2}{\rm Log}\left(\sum_{k=0}^\infty {N_3+k-1\choose k}{N_2-N_1+k\choose k}
k! \left({t\over N}\right)^k \right)\cr}} 
where the term $(1-x)^{-N_3}$ is only meant as a formal power series of $x$.
Again, the final result for the free energy is obtained by taking $f_n\vert_{n=N_1}$.

It is quite interesting to compare this result for the all-genus free energy
of the vertex-tricolored triangulations to that of \DEG, reading
with the present notations:
\eqn\readdeg{\eqalign{
\phi_{n+1}+\phi_{n-1}-2 \phi_n&={1\over N^2}{\rm Log}
\left(1+{Nt\over n}(t\partial_t)^2 \phi_n\right)\cr
\phi_0&=0\cr
\phi_1&={1\over N^2}{\rm Log}\left(\sum_{k=0}^\infty {N_3+k-1\choose k}{N_2+k-1\choose k}
k! \left({t\over N}\right)^k \right)\cr}}
and the final result for the free energy is obtained by taking $\phi_n\vert_{n=N_1}$ as before.
That \recufn\ and \newinit\ lead to the same result as \readdeg, namely
that $\phi_{N_1}=f_{N_1}$, is a highly non-trivial fact.

\subsec{Planar limit}

The planar limit of the free energy is easily obtained from the results of
Sect.4.2, again by 
first having infinitely many $g_k$'s in the potential \genepot, 
and by finally taking $g_k=x_3$ for all $k\geq 1$. 
Using the generating functions \genphik\ \genkphi\ of the $\varphi_k$'s and $\phi_k$'s at $z=1$,
and substituting the result into the extended versions of \readfin\ and \recasthem,
we get
\eqn\lutpresq{\eqalign{
E(t)&= {U_1U_2\over t^2}-{x_3\over 2 t}\bigg({1-U_1-U_2\over 
\sqrt{(1-U_1-U_2)^2-4 U_1U_2}} -1\bigg) -x_1 x_2 \cr 
t x_1&= U_1\bigg({1\over t} -{x_3\over 2} \big({1+U_1-U_2\over 
\sqrt{(1-U_1-U_2)^2-4 U_1U_2}} -1\big) \bigg)\cr
t x_2&= U_2\bigg({1\over t} -{x_3\over 2} \big({1+U_2-U_1\over  
\sqrt{(1-U_1-U_2)^2-4 U_1U_2}} -1\big) \bigg)\cr}} 
We then use the following change of variables to get rid of the square roots:
setting $U_1=V_1(1-V_2)$ and $U_2=V_2(1-V_1)$ indeed transforms
$(1-U_1-U_2)^2-4 U_1U_2$ into a square $(1-V_1-V_2)^2$. This results in
\eqn\lutinter{ \eqalign{
E(t)&={V_1V_2\over t^2}\big((1-V_1)(1-V_2)-{tx_3\over 1-V_1-V_2}\big) -x_1x_2\cr
t x_1&=V_1\big(1-V_2 -{t x_3\over 1-V_1-V_2}\big) \cr
t x_2&=V_2\big(1-V_1 -{t x_3\over 1-V_1-V_2}\big) \cr}}
Introducing finally a third variable $V_3=t x_3/(1-V_1-V_2)$, we finally recover
the result of \DEG: 
\eqn\lutfin{\eqalign{
E(t)&={V_1V_2V_3\over t^2}(1-V_1-V_2-V_3) \cr
tx_1&=V_1(1-V_2-V_3)\cr
tx_2&=V_2(1-V_1-V_3)\cr
tx_3&=V_3(1-V_1-V_2)\cr}}
where $V_i=tx_i+O(t^2)$ are to be expanded in power series of $t$ and substituted into
the expression for $E$.
This result was given a purely combinatorial proof in Ref. \BDE, by suitably mapping
marked tricolored triangulations to colored trees generated by the $V_i$'s.

\newsec{Rectangular multi-matrix models for new random lattice statistical models}

In this paper, we have shown how complex rectangular one-matrix models 
could be used to generate and count various types
of face-colored graphs. 
The methods of solution presented are adapted from the Hermitian matrix case,
and lead in particular to closed formulas for the planar generating functions,
expressed in terms of tree-like generating functions, thus providing an
alternative proof of some recent combinatorial results based on
some bijections between rooted colored planar graphs and some suitably colored trees.

The standard methods of Hermitian matrix models can also be adapted to the
rectangular multi-matrix case, by use of the Harish-Chandra-Itzykson-Zuber 
formula \HCIZ, namely that 
\eqn\izfor{ \int_{U(N)} dU e^{{\rm Tr}(aUbU^\dagger)} 
\propto {\det\big[e^{a_ib_j}\big]_{1\leq i,j\leq N} \over \Delta(a)\Delta(b)}}
where the integral is performed over unitary $N\times N$ matrices and
$a,b$ are some diagonal matrices of same size. An interesting generalization 
of this formula was also derived in Refs. \BROzub\ \VERzub\ \WADzub\ 
for rectangular matrices and reads:
\eqn\verba{ \int_{U(N_1)} dU \int_{U(N_2)} dV e^{{\rm Re}\, {\rm Tr}(UAV^\dagger 
B^\dagger)}
\propto 
{\det \big[I_{N_2-N_1}(a_ib_j)
\big]_{1\leq i,j\leq N_1} \over 
\Delta(a^2)\Delta(b^2) 
\prod_{i=1}^{N_1} (a_i b_i)^{N_2-N_1} 
} }
where $A$ and $B$ are rectangular matrices of size $N_1\times N_2$ 
(with $N_1\leq N_2$), $a$ and $b$ are the positive square roots
of the diagonalized versions of $AA^\dagger$ and $B^\dagger B$ respectively, 
and $I_n$ are the modified Bessel functions.
Rectangular multi-matrix integrals with chain-like interactions are 
expected to generate more involved face-colored graphs, and can be reduced to 
eigenvalue integrals by use of either \izfor\ or \verba\ depending on the form
of the interaction terms ($AA^\dagger BB^\dagger$ or $AB^\dagger+BA^\dagger$). 

We give below two series of examples of new rectangular multi-matrix models, 
whether solvable or not, describing random lattice statistical models.

\subsec{IRF Models on random tessellations}

Hermitian matrix models are well adapted for generating configurations of
{\it vertex} models, as the terms in the potential are directly related
to vertex configurations.
Here we concentrate on so-called Interaction-Round-a-Face (IRF) statistical models,
originally defined on fixed lattices by assigning height variables to the vertices
of the lattice (say a map $i\to h(i)$ from vertices to a target space $T$),
and attaching to each configuration of these heights a Boltzmann weight, itself
a product of local Boltzmann weights depending only on the heights {\it around each 
elementary face} of the lattice. A tremendous amount of work has been done for
solving fixed lattice IRF models, in relation with integrability. 
When defined on a random lattice, these require that the faces of the
random tessellation replacing the lattice keep their original form, so that the
local Boltzmann weights remain the same. A nice global picture for a
particular class of random lattice IRF models (the so-called Restricted Solid On Solid models
based on the A,D,E Dynkin diagrams)
was obtained in \IK, using first the correspondence
between these models and loop gases in the planar case, and then using
a method of surgery to generate higher genus contributions.

\fig{The four vertices of the matrix model describing
the general two-state IRF model on a random triangulation. These vertices are dual to
the faces of the triangulation. According to the nature of the matrix indices running over
the lines (the small circled numbers $1,2$ stand for the ranges $j=1,2,...,N_1$ or $N_2$
respectively), the faces (i.e. the vertices of the dual triangulation) may be colored
accordingly with $1,2$ (the large circled numbers $1,2$). The corresponding Boltzmann weights
$g_{1,2,3,4}$ are indicated.}{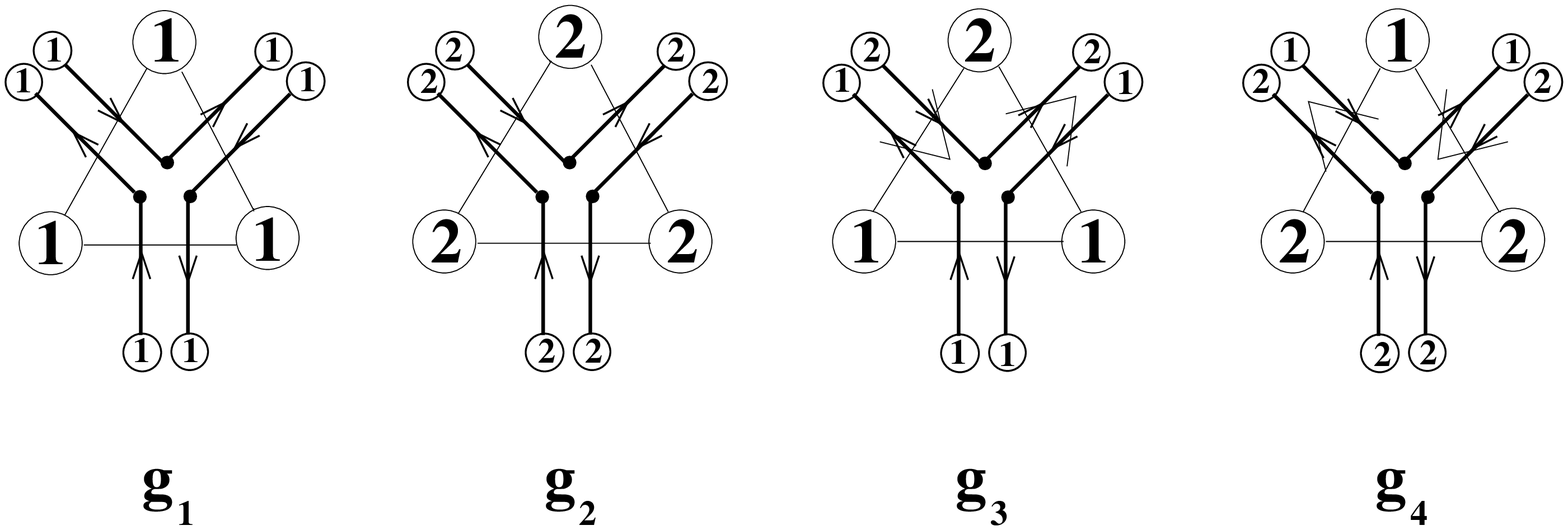}{10.cm}
\figlabel\boltz

Let us show how rectangular matrix models naturally generate general IRF models.
The trick is to use a dual picture of IRF models, in which the height variable is
a {\it face} variable (say a color) defined on the {\it irregular} faces of a 
random lattice with regular vertices. The particular choice of potential fixes
the corresponding vertex Boltzmann weights, which dually translate into the
local face boltzmann weights of the IRF model.
For simplicity, we start with IRF models on random triangulations. We use a rectangular
matrix model to generate the dual (trivalent) graphs, for which the original height variable
$h(i)$ simply becomes the color of the face $i$, itself indicated by the range of matrix
indices running along the loop bordering the face $j=1,2,...,N_i$. 
For instance, in the case of a target with two elements, say $T=\{1,2\}$,
we must assign colors $1,2$ to the faces of random trivalent graphs, with specific
Boltzmann weights $g_{1,2,3,4}$ depending on the three colors around each vertex, as shown 
in Fig.\boltz. 
This is readily realized by a triple integral over two Hermitian and one rectangular
matrices $A$ of size $N_1\times N_1$, $B$ of size $N_2\times N_2$ and $C$ of size $N_1\times N_2$:
\eqn\deficoz{\eqalign{
Z&={\int dA dB dC e^{-N{\rm Tr}(V(A,B,C))}\over \int dA dB e^{-N{\rm Tr}(V_0(A,B,C))}}\cr 
V_0(A,B,C)&={A^2\over 2}+{B^2\over 2}+CC^\dagger\cr
V(A,B,C)&=V_0(A,B,C) -g_1 {A^3\over 3} -g_2{B^3\over 3}-g_3ACC^\dagger
-g_4 B C^\dagger C\cr}}
Note that the potentials $V$ and $V_0$ only make sense here within a trace, as $A$ and $B$ have
different sizes.
As before, the logarithm of $Z$ in \deficoz\ generates the configurations of the most general
$2$-state IRF model on triangulations of arbitrary genus $h$, with a weight $N^{2-2h}$,
provided we have set $N_i=x_i N$.
Particular cases of this model include: 
(i) the triangular Ising model, with $N_1=e^HN$, $N_2=e^{-H}N$,
$g_1=g_2=g e^{3K/2}$, $g_3=g_4=g e^{-K/2}$, where $H$ is the magnetic
field,  $K=J/(k_BT)$ the normalized spin coupling and $g$ the cosmological constant
and  (ii) Baxters's triangular three-spin model
\BAXBOOK\ with an additional magnetic field $H$, with $N_1=e^HN$, $N_2=e^{-H}N$, $g_1=g_4=g e^{K}$,
$g_2=g_3=g e^{-K}$. 
As the matrix $C$ occurs only in quadratic terms in the potential $V$ of \deficoz, 
it may be integrated explicitly. Indeed, writing this quadratic form
as $\sum_{ijkl} C_{ij}{\bar C}_{lk} (\delta_{jk}\delta_{li}-g_3 \delta_{jk} A_{li} 
-g_4 B_{jk} \delta_{li})$, the integration over $C$ allows to rewrite the model
as a two-Hermitian matrix model, with matrices $A$ and $B$ of different sizes
$N_1$ and $N_2$, and with potentials
\eqn\finpothree{\eqalign{
V_0(A,B)&= {A^2\over 2}+{B^2\over 2} \cr
V(A,B)&=V_0(A,B)+{1\over N}{\rm Log}\left(I_{N_1}\otimes I_{N_2}-g_3 
A\otimes I_{N_2}-g_4 I_{N_1}\otimes B\right)-g_1 {A^3\over 3} -g_2{B^3\over 3}\cr}}
where the subscript of the identity refers to the size of the matrix, and again the potentials
only make sense within a trace as $A$ and $B$ have different sizes.
This extends easily to the case of $k$-state IRF models. Note that in the 
particular case of $A,D,E$-type
targets, where states of adjacent vertices of the triangulation may either be of the same
color or of neighboring ones on a vertex-colored target graph of $A,D,E$-type, the rectangular matrices
can all be integrated out, leaving us with a slight generalization of Ref. \KKO\ in which matrices
may have different sizes, which has the effect of introducing a ``magnetic field" coupled
to colors.  
Let us also mention the square version of \deficoz\ for completeness. It generates the most general
two-state IRF model on random quadrangulations:
\eqn\defcoq{\eqalign{
Z&={\int dA dB dC e^{-N{\rm Tr}(V(A,B,C))}\over \int dA dB e^{-N{\rm Tr}(V_0(A,B,C))}}\cr 
V_0(A,B,C)&={A^2\over 2}+{B^2\over 2}+CC^\dagger\cr
V(A,B,C)&=V_0(A,B,C) -g_1 {A^4\over 4} -g_2{B^4\over 4}-g_3A^2CC^\dagger \cr
&-g_4 B^2 C^\dagger C-g_5 ACBC^\dagger -g_6{(CC^\dagger)^2\over 2}\cr}}
Note that if $g_6\neq 0$, the rectangular matrix $C$ can no longer 
be integrated out as in the previous case.

It is now clear that increasing the number of matrices will allow for exploring larger target
spaces $T$.
One other case of interest is the straightforward generalization of the 
vertex-tricolored triangulations of Sect.5 into random $k$-gonal tessellations with
vertices colored $1,2,...,k$ cyclically around each face, clockwise 
or counterclockwise in alternance, weighted by $x_i$ per vertex of color $i$
and $t$ per edge. The planar free energy for this problem was derived
recently in \BDE\ and is also related to work on constellations \BMS.
The $k$ colors require the introduction of $k$ rectangular matrices $A_1,A_2,...,A_k$ of
respective sizes $N_1\times N_2,N_2\times N_3,...,N_k\times N_1$, while the coloring
rule is encoded in two (clockwise and counterclockwise) vertices $A_1A_2...A_k$
and its adjoint. The partition function for this problem reads
\eqn\pfkgo{\eqalign{
Z&={\int \prod_{i=1}^k dA_i e^{-N{\rm Tr}(V(A_1,...,A_k))} \over
\int \prod_{i=1}^k dA_i e^{-N{\rm Tr}(V_0(A_1,...,A_k))}}\cr
V_0&=\sum_{i=1}^k {A_iA_i^\dagger\over t}\cr 
V&=V_0-g(A_1A_2...A_k+A_k^\dagger A_{k-1}^\dagger ...A_1^\dagger )\cr}}
Taking again $N_i=x_iN$ yields the desired all-genus free energy
$F={\rm Log}\, Z$.
The particularly simple form of the planar free energy found in \BDE\ suggests
that this model should be solvable.

\subsec{Hard objects on random tessellations}

Another class of interesting models is that of mutually excluding (so-called hard) particles.
On a fixed lattice, vertices may be empty or occupied by a particle, with the constraint
that no two particles may occupy adjacent vertices, and one moreover attaches an activity
$z$ per particle. The case of the triangular lattice was solved by Baxter \BAXHH\
under the name of hard hexagons (the area excluded by a particle on the triangular
lattice is the hexagon formed by its six adjacent vertices). Neither the square nor
the hexagonal lattice cases were solved exactly yet although much
is known on them \BAX\ \BAXHS, in particular thanks to the powerful corner
transfer matrix method developed by Baxter \BAXBOOK. 
As far as critical properties are concerned, all these models are expected 
to undergo a crystallization transition at some positive value $z=z_c$ of the activity,
between a fluid low-density phase and a crystalline high-density phase, in which particles
tend to maximally occupy a sublattice of the lattice. Remarkably, this clearly distinguishes the
triangular lattice case from the two others, as there are {\it three} candidate sublattices
for the crystal as opposed to only two in the square and hexagonal cases. As a consequence,
the hard hexagon model lies at the crystallization transition in the
universality class of the critical three-state Potts model (unitary CFT
with central charge $c=4/5$), while in the two other cases, the expected
universality class is that of the critical Ising model (unitary CFT
with central charge $c=1/2$).

\fig{The three vertices used via matrix model to generate the dual graphs to the configurations
of the hard particle model on random quadrangulations. We have indicated the nature of the
matrix index lines by a small circle (white for indices ranging from $1$ to $N$,
grey for indices ranging from $1$ to $P=zN$). We have also represented in thin lines
the squares dual to the vertices, and the corresponding status of their 
(empty or occupied) vertices by large (white or grey) circles. 
The Boltzmann weights $g_{1,2,3}$ for the three 
situations are also indicated.}{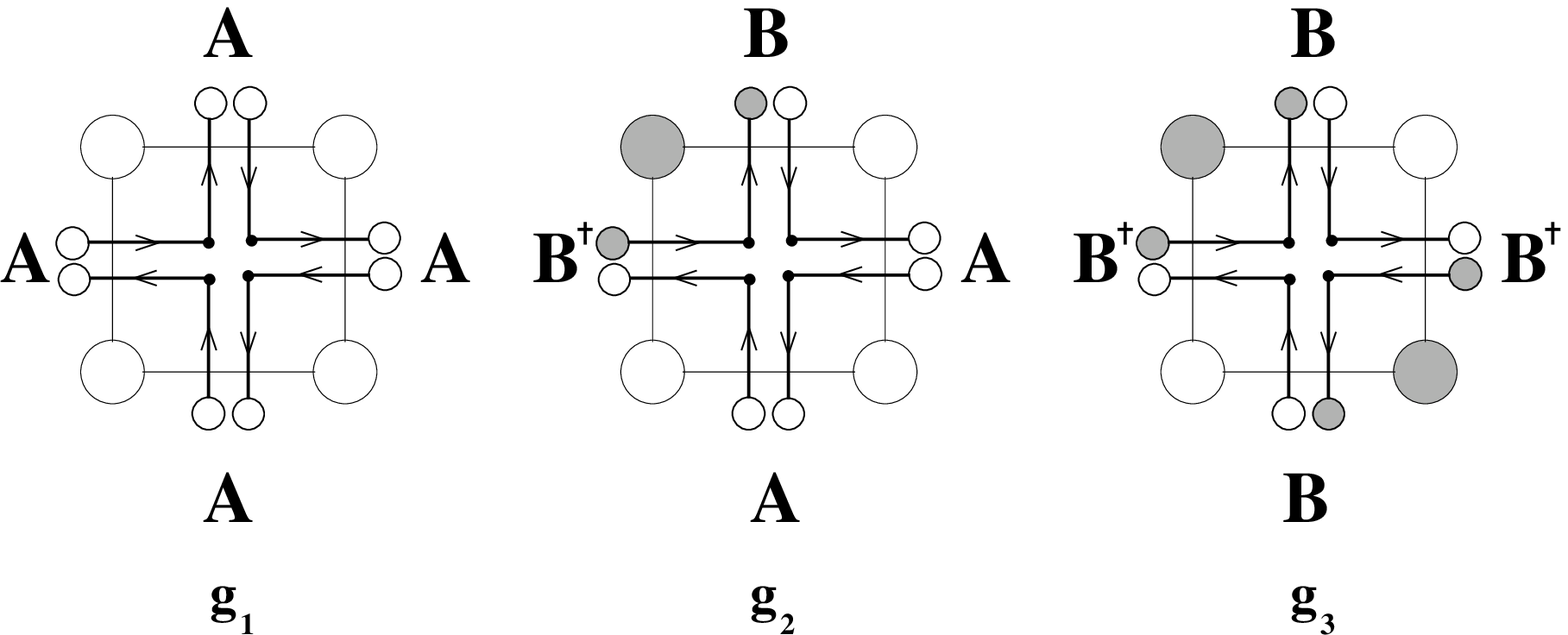}{10.cm}
\figlabel\hardsq

\fig{A typical configuration of the hard-particle model on the faces
of a random tetravalent planar graph, as obtained from some particular mixed
square and rectangular matrix integral. Occupied faces are represented in grey,
while empty ones are white. These correspond respectively to indices ranging
from $1$ to $P$ and $N$ on the oriented line bordering them. When summed
over, these result in a weight $N$ per empty face and $P$ per occupied one.
By duality, this graph is nothing but a random quadrangulation with hard-particles
at some of its vertices.}{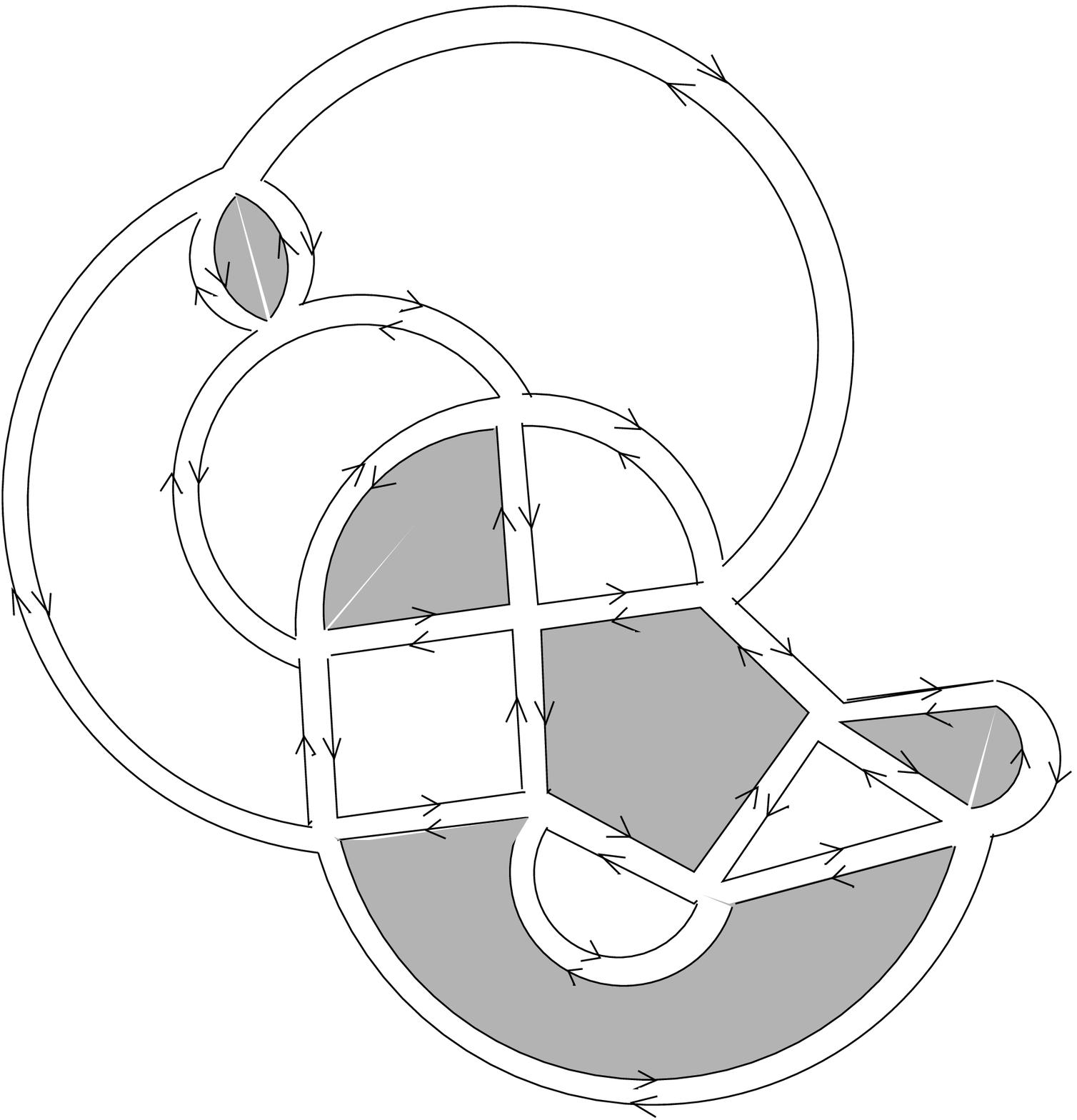}{6.cm}
\figlabel\sqhard
 
Recently, it was shown in \BDFG\ that 
the expected critical properties for square and hexagonal lattices
are correctly reproduced on random vertex-bicolorable lattices, which still allow
for the existence of two crystalline states. These involved however lattices
with faces of arbitrary valences. Rectangular matrices can be used to generate
analogous models, but with regular faces, the randomness being concentrated into
the arbitrary valences of vertices.
For the case of hard squares, we must generate random quadrangulations with empty and occupied
vertices according to the exclusion rule. Note that the vertices are naturally bicolored,
and that we therefore expect to reproduce the correct critical behavior of hard squares, as
we will always have two natural maximally occupied crystalline states.
This requires simply two matrices, one square Hermitian $A$ of size $N\times N$ and one
rectangular $B$ of size $N\times P$, whose integral will generate the duals of the decorated
quadrangulations. 
Indeed, using the three vertices displayed in Fig. \hardsq, we may attach to the faces of the 
tetravalent graph they form once connected two colors say white and grey (empty or occupied) 
according to whether the running matrix index around the face ranges from 
$1$ to $N$ or $P$, as illustrated in Fig. \sqhard.
The allowed vertices of Fig. \hardsq\ carry the information on the exclusion rule. 
The partition function for the hard particle model on
random quadrangulations then reads: 
\eqn\parhs{\eqalign{
Z&={\int dA dB e^{-N{\rm Tr}(V(A,B))} \over \int dA dB e^{-N{\rm Tr}(V_0(A,B))}}\cr
V_0(A,B)&= {A^2\over 2} +B B^\dagger\cr
V(A,B)&=V_0(A,B) - g_1 {A^4 \over 4} -g_2 A^2 B B^\dagger -g_3 {(BB^\dagger )^2\over 2}\cr}}
Indeed, upon setting $P=N z$, the logarithm of $Z$ of \parhs\ generates the all-genus connected
quadrangulations with hard particles on their vertices, and a weight $N^{2-2h}$, $h$ the genus, 
a weight $z$ per particle and extra weights $g_1$ per empty square, $g_2$ per singly occupied
square and $g_3$ per doubly occupied square (which can be interpreted as a diagonal interaction),
and finally the usual inverse symmetry factor. The integral \parhs\ may be transformed into one
over two Hermitian matrices as follows. We first form the square
\eqn\forsq{ -g_2 A^2 B B^\dagger -g_3 {(BB^\dagger )^2\over 2}=-{g_3\over 2}
(BB^\dagger +{g_2\over g_3}A^2)^2+{g_2^2\over 2 g_3} A^4 }
which we represent by the use of a $N\times N$ Hermitian matrix $M$ as
\eqn\sqrep{e^{-N{\rm Tr}(-{g_3\over 2}(BB^\dagger +{g_2\over g_3}A^2)^2)}=
{\int dM e^{-N {\rm Tr}({M^2\over 2}-\sqrt{g_3} M (BB^\dagger +{g_2\over g_3}A^2))}
\over \int dM e^{-N{\rm Tr}({M^2\over 2})} } } 
Integrating over $B$ is readily done using \genint,
and now yields an additional term $z {\rm Log}(1-\sqrt{g_3}M)$ in the
potential. We therefore end up with an integral over the two Hermitian matrices $A,M$ of size
$N\times N$, with potentials
\eqn\potwomat{\eqalign{
V_0&={A^2\over 2}+{M^2\over 2} \cr
V&=V_0-{g_2\over \sqrt{g_3}} M A^2 -\left( g_1-2{g_2^2\over g_3}\right){A^4 \over 4}
+z {\rm Log}(1-\sqrt{g_3}M)\cr}}
This is a particular generalization of the $O(n=1)$ model coupled to discrete 2D quantum gravity.
When $g_1 g_3=2 g_2^2$, it indeed reduces to a particular $O(n=1)$ model, with a logarithmic
potential. We expect its (multi-) critical point to be that of the critical Ising model coupled
to 2D quantum gravity. 

\fig{The allowed vertices for the random lattice version of the hard hexagon
model, via rectangular matrix integral.
The small white (resp. grey) circled numbers $i=1,2,3$
refer to matrix indices ranging from $1$ to $N_i$ (resp. $P_i=z N_i$), while the
large white (resp. grey) ones refer to the color of the empty (resp. occupied)
corresponding face or vertex of the dual triangulation. We have
indicated the nine different rectangular matrices needed for the model.}{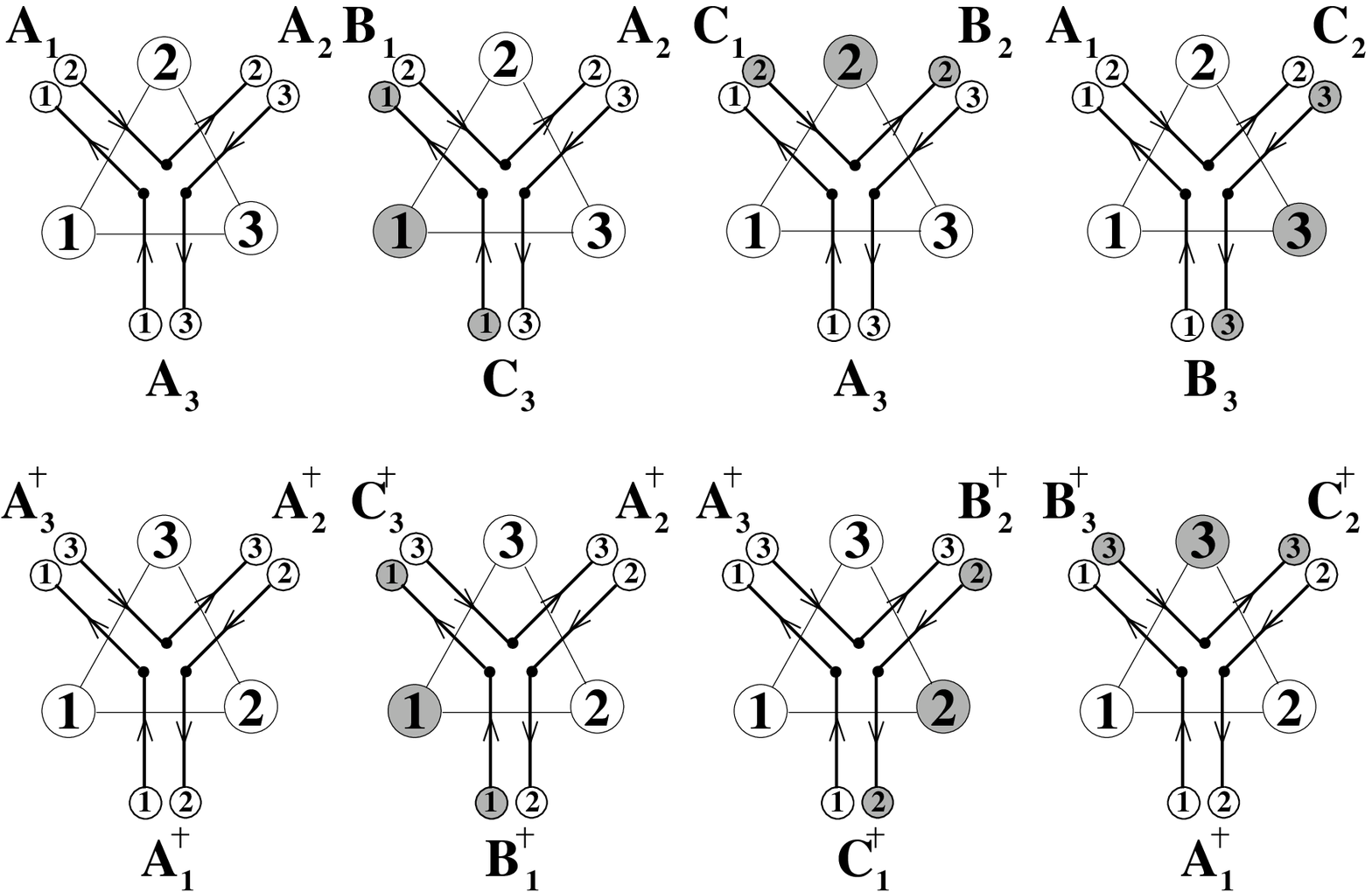}{10.cm}
\figlabel\boltri

We may now also write a candidate model for the random surface version of hard hexagons,
by considering the hard particle gas on a vertex-tricolored triangulation. Indeed, 
we expect just like in \BDFG\ that preserving the tricolorability of 
the triangulation will allow
for the existence of three competing crystalline states, eventually leading to
the critical transition of the three-state Potts model on a random surface. 
The idea is simply to attach a different color to occupied colored faces, say
by replacing the range of matrix indices $j=1,2,...,N_i$
(for an empty face of color $i$) by $j=1,2,...,P_i=z N_i$ (for an occupied face
of color $i$). The exclusion rule is then enforced through the allowed vertex 
configurations of Fig. \boltri. 
The model reads as follows:
\eqn\readhard{\eqalign{
Z&={\int \prod_{i=1}^3 dA_i dB_i dC_i 
e^{-N{\rm Tr}(V(\{A_i,B_i,C_i\}))}\over \int \prod_{i=1}^3 dA_i dB_i dC_i 
e^{-N{\rm Tr}(V_0(\{A_i,B_i,C_i\}))} } \cr
V_0&=\sum_{i=1}^3 A_iA_i^\dagger +B_iB_i^\dagger+C_iC_i^\dagger \cr
V&=V_0-g (A_1A_2A_3+B_1A_2C_3+C_1B_2A_3+A_1C_2B_3+\ {\rm h.c.}) \cr}}
where h.c. denotes the hermitian conjugate of the sum of four terms
(the total of eight terms corresponds to the vertices of Fig. \boltri), and
where the nine rectangular matrices have respective sizes
\eqn\sizesmat{\matrix{ A_1:N_1\times N_2 & A_2:N_2\times N_3 & A_3:N_3\times N_1\cr
B_1:zN_1\times N_2 & B_2: zN_2\times N_3 & B_3: zN_3\times N_1\cr
C_1:N_1\times zN_2 & C_2:N_2\times zN_3 & C_3:N_3\times zN_1\cr}}
The logarithm of $Z$ \readhard\ generates the configurations of the hard-particle
model on vertex-tricolored random triangulations of genus $h$, with a weight $N^{2-2h}$
provided we take $N_i=x_i N$, $i=1,2,3$, and also with a weight $z$ per particle
and $x_i$ per face of color $i$.
The partition function $Z$ of \readhard\ may be further simplified by explicitly integrating
out (i) the $C_i$, $i=1,2,3$, which results in new terms 
$-g^2(B_1A_2A_2^\dagger B_1^\dagger+B_2A_3A_3^\dagger B_2^\dagger+B_3A_1A_1^\dagger B_3^\dagger)$
for the potential and finally (ii) the $B_i$, $i=1,2,3$, by use of the formula \genint. 
The final result is an integral over 
the three rectangular matrices $A_i$:
\eqn\fintrihard{\eqalign{
Z&={\int dA_1dA_2dA_3 e^{-N{\rm Tr}(V(A_1,A_2,A_3))} \over
\int dA_1dA_2dA_3 e^{-N{\rm Tr}(V_0(A_1,A_2,A_3))}}\cr
V_0&=A_1A_1^\dagger+A_2A_2^\dagger+A_3A_3^\dagger \cr
V&=V_0-g(A_1A_2A_3+A_3^\dagger A_2^\dagger A_1^\dagger)
+zx_1{\rm Log}(1-g^2A_2A_2^\dagger)\cr
&+zx_2{\rm Log}(1-g^2A_3A_3^\dagger)+zx_3{\rm Log}(1-g^2A_1A_1^\dagger)\cr}}

This is easily generalized to the hard-particle model on the vertices of
a cyclically $k$-colored random $k$-gonal tessellation (with alternating
clockwise and counterclockwise order), with the additional constraint that
a given $k$-gon have its vertices occupied by a total of at most one particle. 
Such a model naturally gives rise to $k$
distinct crystalline states. The corresponding rectangular multi-matrix model
is reduced in an analogous way to an integral over $k$ matrices $A_1$, $A_2$,... $A_k$
with sizes $N_1\times N_2$, $N_2\times N_3$, ... $N_k\times N_1$, and with
potentials
\eqn\potfinal{\eqalign{
V_0&=\sum_{i=1}^k A_i A_i^\dagger \cr
V&=V_0-g(A_1A_2...A_k+A_k^\dagger A_{k-1}^\dagger ...A_1^\dagger )
+\sum_{i=1}^k z x_i 
{\rm Log}\bigg(1-g^2 \big(\prod_{j=1\atop j\neq i-1,i-2}^k A_j\big) \times {\rm h.c.}\bigg)\cr}} 
where we have extended the definition of indices in such a way that $0\equiv k$
and $-1\equiv k-1$ (namely when $i=1$ the terms $A_{k-1}$ and $A_k$ must be omitted in the product,
while when $i=2$ $A_1$ and $A_k$ must be omitted). 
In the case $k=4$ of hard particles on cyclically vertex-$4$-colored quadrangulations
with at most one particle per square, we expect the model to reach a crystallization
transition in the universality class of the random lattice 
critical $4$-state Potts model (CFT with central charge $c=1$).
Note that this point is not reachable within the model \parhs\
of hard-particles on plain quadrangulations with $g_3=0$ (i.e. with at most
one particle per square), for which no natural maximally
occupied crystalline state will survive.  
More precisely, taking $g_3=0$ in \parhs\ allows to integrate over
the rectangular matrix $B$, leaving us with a one-Hermitian matrix model with
potentials
\eqn\onehermi{
V_0(A)={A^2\over 2}, \qquad V(A)={A^2\over 2}+z{\rm Log}(1-g_2 A^2)-g_1{A^4\over 4} } 
and it is a simple exercise to check that the only multicritical points
of this model all correspond to negative critical values of $z$
and therefore lie in the universality class of the random lattice 
Lee-Yang edge singularity (non-unitary CFT with $c=-22/5$), meaning 
that the (unitary) crystallization transition has
been wiped out by the sum over random quadrangulations.

\bigskip

\noindent{\bf Acknowledgements}

This work originated from discussions with J. Bouttier and E. Guitter. 
We also thank J.-B. Zuber for useful discussions and
especially G. Akemann for a critical reading of the manuscript and 
some help with the references.  

\appendix{A}{Planar Eulerian Graphs from the one-Hermitian Matrix Model}

We wish to compute the large $N$ limit of the function $Z$ \hermat,
in the even potential case, namely with $V(x)=\sum_{i=1}^J g_{2i}{x^{2i}\over 2i}$.
The cutoff degree $2J$ is purely artificial, but dealing with polynomials
greatly simplifies the calculations. As we are only interested in formal power
series, we will simply take $J\to\infty$ in the final result.
To compute $Z$ of \hermat, we first reduce the matrix integral 
to an eigenvalue integral by changing variables $M\to (U,m)$, where $M=UmU^\dagger$,
where $m=$diag$(m_1,...,m_N)$
and $U\in U(N)/U(1)^N$ are some angular variables. The Jacobian of the transformation 
is well-known to be the Vandermonde determinant squared $\Delta(m)^2$, with
$\Delta(m)=\prod_{i<j} (m_j-m_i)$. One is therefore left with the computation
of 
\eqn\ofZ{\eqalign{ Z&= {Z(V)\over Z(V_0)} \cr
Z(V)&=\int dm e^{-N^2 S(m,V)} \cr
S(m,V)&= {1\over N} \sum_{i=1}^N V(m_i) 
-{1\over N^2}\sum_{1\leq i\neq j\leq N}{\rm Log}|m_i-m_j| \cr
V(x)&={x^2 \over 2t} -\sum_{i=1}^J g_{2i} {x^{2i}\over 2i} \cr
V_0(x)&={x^2 \over 2t}\cr}} 
We now evaluate separately the numerator and denominator in the large
$N$ limit, where the corresponding integrals are dominated by the
saddle-point of the action functional $S$, namely
the $m_i$'s such that $\partial S/\partial m_i=0$, hence
\eqn\sapoh{ V'(m_i)={2\over N} \sum_{j\neq i} {1\over m_i-m_j}}  
This leads to a quadratic equation for the planar resolvent
\eqn\resoh{ \omega(z)=\lim_{N\to \infty} {1\over N}\sum_{i=1}^N {1\over z-m_i} }
evaluated on the solution $m$ to \sapoh, and related to the limiting density of eigenvalues
\eqn\lidensi{ \rho(z)= \lim_{N\to \infty} {1\over N}\sum_{i=1}^N \delta(z-m_i)}
through $\rho(z)=(\omega(z-i0)-\omega(z+i0))/(2i\pi)$.
The solution takes the form
\eqn\solresh{ \omega(z)={1\over 2} \big( V'(z) +\sqrt{V'(z)^2-4 P(z)} \big) }
for say $g_{2I}>0$, 
where $P(z)=\lim_{N\to \infty}{1\over N}\sum_{i=1}^N (V'(z)-V'(m_i))/(z-m_i)$
is a polymomial of degree $M-2$. $\omega(z)$ is further fixed by its large
$z$ behavior $\omega(z)\sim 1/z$ and by also demanding that it have
only a one-cut square root singularity in the complex plane. This amounts
to saying that $V'^2-4 P$ has $M-2$ double zeros and two single ones say $a,b$, so
that finally $\omega(z)=(V'(z)+Q(z)\sqrt{(z-a)(z-b)})/2$, $Q$ a polynomial
of degree $M-2$. 
Moreover, the potential $V(x)$ being an even function of $x$, it is clear that 
the density of eigenvalues $\rho$ is also an even function, and therefore $b=-a$.
The (even) polynomial $Q$ is fixed by requiring that the large $z$ expansion of $\omega(z)$
have no positive powers of $z$. This gives
\eqn\calQh{ Q(z)=-\left[ {V'(z)\over \sqrt{z^2-a^2}} \right]_+ }
where we must expand the bracket at large $z$, and the subscript $+$ indicates
that one only retains the polynomial part of the expansion. Finally, the value
of $a$ is determined as follows.
Writing $R(z)=-V'(z)/\sqrt{z^2-a^2}$ as a series expansion at $z=\infty$,
we also get that $\omega(z)=-{1\over 2} \big[ R(z) \big]_- \sqrt{z^2-a^2}$,
where the subscript $-$ means that we only retain the negative powers of 
$z$ in the expansion. As $V'$ is odd, the constant term $z^0$ vanishes and we
must simply express that the residue ($1/z$ term) of $\omega(z)$
is equal to $1$, giving the relation $R_{-2}=-2$,
where the subscript indicates that we only retain the corresponding power of 
$z$ term in the expansion of $R$.
Explicitly writing
\eqn\expansqrt{ {1\over \sqrt{z^2-a^2}}=\sum_{k\geq 0} {I_k \over z^{2k+1}}=\sum_{k\geq 0}
{2k \choose k} {\big({a^2\over 4}\big)^k\over z^{2k+1}}} 
and further defining $I_k=0$ for $k<0$, 
we get
\eqn\expliR{ R(z)= \sum_{k=1-J}^{\infty} {1\over z^{2k}} \big( -{I_{k}\over t}+
\sum_{i=1}^J g_{2i} I_{i+k-1} \big)  }
which, applied to the relation $R_{-2}=-2$, finally leads to the equation determining $a$:
\eqn\couplab{
{a^2 \over 4t} -\sum_{i=1}^J g_{2i} {2i-1 \choose i} 
\left({a^2\over 4}\right)^i =1 }
The single zero $a$ of the eigenvalue density is the unique formal series solution to \couplab\
that starts as $a=2 \sqrt{t}(1+O(t,\{g_{2i}\}))$ for small positive $t$ and $g$'s.
 
Finally the planar free energy reads
$f(V)=\lim_{N\to \infty} (1/N^2){\rm Log}Z(V)=-S(m,V)$ taken at the 
saddle-point solution. We also have
\eqn\calfs{ t{df\over dt}= 
-\sum_{i=1}^N {\partial S \over \partial m_i}\, t {dm_i\over dt}
-t{\partial S\over \partial t} = \lim_{N\to \infty} {1\over N}\sum_{i=1}^N {m_i^2\over 2t}=
{\omega_2\over 2t} }
where we have expanded the resolvent as $\omega(z)=1/z+\omega_1/z^2+\omega_2/z^3+O(1/z^4)$
at large $z$. We may therefore identify $tdf/dt$ with  
\eqn\identiwi {t {df\over dt}={1\over 8t}\big(-2 R_{-4} -2a^2\big)}
For the Gaussian potential $V_0$, the equation \couplab\ reduces to
$a^2=4t$, so that $a=2\sqrt{t}$, $R_{-4}=-I_2/t=-6 t$ and $tdf_0/dt=1/2$.
Finally, the generating function for arbitrary Eulerian planar graphs, 
with a weight $t$ per edge, and $g_{2i}$ per $2i$-valent vertex 
and with a marked oriented edge (with a net effect of $2td/dt$) reads
\eqn\Efinh{\eqalign{
E(t)&=2t{df\over dt}-2t {df_0\over dt}={1\over 2t} \big(-
R_{-4} -a^2-2 t\big)\cr
&={1\over 2t} \big({I_2\over t} 
-\sum_{i=1}^M g_{2i} I_{i+1} -a^2-2 t\big)\cr} }
Let us finally introduce the function
$U(t)=a^2/4$.
With this definition, eqn. \couplab\ reduces to
\eqn\treegenh{ {U \over t} -\sum_{i=1}^J g_{2i} {2i-1\choose i} U^i=1  }
while \Efinh\ gives
\eqn\lutfh{\eqalign{
E(t)&={3U^2\over t^2}-{2U\over t}-1-{1\over t}\sum_{i=1}^J
g_{2i}{2i+1\choose i}U^{i+1} \cr
&={U\over t}\big(1-\sum_{i=2}^J g_{2i} {2i-1\choose i-2}
U^i\big)-1\cr}}
It is also interesting to compute the following derivative of $E$:
\eqn\derivE{ {d \over dt} (tE(t)) = {U^2\over t^2}-1 }
where we have used \treegenh\ to eliminate $dU/dt$.
This relation, due to Bender and Canfield \BC, 
was given a combinatorial interpretation in Ref. \SCH,
by establishing a bijection between suitably marked Eulerian graphs
and pairs of trees generated by the function $U(t)$.
Note that in this approach the Gaussian free energy $f_0$ is not subtracted 
from the free energy $f$, resulting in a shift $E\to E+1$.

\listrefs
\end